\documentclass[]{aa}  

\usepackage{graphicx}
\usepackage{txfonts}

\newcommand{\Gaia}{{\it Gaia}}

\usepackage[]{hyperref}
\usepackage{amsmath}
\usepackage{xcolor}

\begin{document}

   \title{Blanco DECam Bulge Survey (BDBS)}

   \subtitle{V. Cleaning the foreground populations from Galactic bulge colour-magnitude diagrams using {\Gaia} EDR3}

   \author{Tommaso Marchetti
          \inst{1},
          Christian I. Johnson
          \inst{2},
          Meridith Joyce
          \inst{2, 3},
          R. Michael Rich
          \inst{4},
          Iulia T. Simion
          \inst{5},
          Michael D. Young
          \inst{6},
          William Clarkson
          \inst{7},
          Catherine A. Pilachowski
          \inst{8},
          Scott Michael
          \inst{8},
          Andrea Kunder
          \inst{9},
          Andreas J. Koch-Hansen
          \inst{10}
          }

\authorrunning{T. Marchetti et al.}
\titlerunning{BDBS V: {\Gaia} Cleaning of Bulge CMDs}

   \institute{European Southern Observatory, Karl-Schwarzschild-Strasse 2, 85748 Garching bei München, Germany\\
              \email{tommaso.marchetti@eso.org}
         \and
             Space Telescope Science Institute, 3700 San Martin Drive, Baltimore, MD 21218, USA
         \and
             Kavli Institute for Theoretical Physics, University of Santa Barbara, California, 93106, USA
         \and
             Department of Physics and Astronomy, University of California Los Angeles, 430 Portola Plaza, Box 951547, Los Angeles, CA 90095-1547, USA
         \and
             Shanghai Key Lab for Astrophysics, Shanghai Normal University, 100 Guilin Road, Shanghai, 200234
         \and
             Indiana University, University Information Technology Services, CIB 2709 E 10th Street, Bloomington, IN 47401 USA
         \and
             Department of Natural Sciences, University of Michigan-Dearborn, 4901 Evergreen Rd. Dearborn, MI 48128, USA
         \and
             Indiana University Department of Astronomy, SW319, 727 E 3rd Street, Bloomington, IN 47405 USA
         \and
             Saint Martin's University, 5000 Abbey Way SE, Lacey, WA 98503, USA
         \and
             Zentrum f\"ur Astronomie der Universitat Heidelberg, Astronomisches Rechen-Institut, Monchhofstr. 12, 69120 Heidelberg, Germany
             }

   \date{Received XXX; accepted YYY}


  \abstract
  {} 
   {The Blanco DECam Bulge Survey (BDBS) has imaged more than 200 square degrees of the southern Galactic bulge, providing photometry in the $ugrizy$ filters for $\sim$ 250 million unique stars. The presence of a strong foreground disk population, along with complex reddening and extreme image crowding, has made it difficult to constrain the presence of young and intermediate age stars in the bulge population.}
   {We employed an accurate cross-match of BDBS with the latest data release (EDR3) from the {\Gaia} mission, matching more than 140 million sources with BDBS photometry and {\Gaia} EDR3 photometry and astrometry. We relied on {\Gaia} EDR3 astrometry, without any photometric selection, to produce clean BDBS bulge colour-magnitude diagrams (CMDs). {\Gaia} parallaxes were used to filter out bright foreground sources, and a Gaussian mixture model fit to Galactic proper motions could identify stars kinematically consistent with bulge membership. We applied this method to $127$ different bulge fields of $1$ deg$^2$ each, with $|\ell| \leq 9.5^\circ$ and $-9.5^\circ \leq b \leq -2.5^\circ$. 
   }
   {The astrometric cleaning procedure removes the majority of blue stars in each field, especially near the Galactic plane, where the ratio of blue to red stars is $\lesssim 10\%$, increasing to values $\sim 20\%$ at higher Galactic latitudes. 
   We rule out the presence of a widespread population of stars younger than 2 Gyr. 
   The vast majority of blue stars brighter than the turnoff belong to the foreground population, according to their measured astrometry. 
   We introduce the distance between the observed red giant branch bump and the red clump as a simple age proxy for the dominant population in the field, and we confirm the picture of a predominantly old bulge.
   Further work is needed to apply the method to estimate ages to fields at higher latitudes, and to model the complex morphology of the Galactic bulge.
   We also produce transverse kinematic maps, recovering expected patterns related to the presence of the bar and of the X-shaped nature of the bulge.
   }
   {}


   \maketitle
%

\section{Introduction}
\label{sec:intro}

The advent of numerous photometric (e.g. 2MASS, \citealt{Skrutskie06}; VVV, \citealt{Minniti10}; OGLE, \citealt{Udalski15}; BDBS, \citealt{Rich20}) and spectroscopic (e.g. BRAVA, \citealt{Rich07, Kunder12}; Gaia-ESO, \citealt{Gilmore12}; ARGOS, \citealt{Freeman13}; GIBS, \citealt{Zoccali14}; APOGEE, \citealt{Majewski17}) ground-based surveys has greatly improved our knowledge on the structure of the Galactic bulge \citep[see][for recent reviews]{Rich13, Babusiaux16, Bland-Hawthorn16, Zoccali16, Barbuy18}. Recently, the data released by the European Space Agency (ESA) satellite {\Gaia} \citep{Gaia16}, including photometry and astrometry for more than 1 billion stars, have revolutionized our understanding of the Milky Way, providing 
essential information complementary to the results of ground-based surveys. In particular, the combined information on precise multi-band photometry, spectroscopy, and astrometry for individual stars in the Galactic bulge has created an invaluable tool for studying its composition, formation, and evolution through cosmic time. This allows for the different stellar populations coexisting along the line of sight towards the central region of our Galaxy to be investigated and disentangled.

The Galactic bulge is known to host a bar \citep[e.g.][]{Binney91, Stanek94}, which forms an angle of $\sim 27^\circ$ with the line of sight from the Sun to the Galactic Centre, and it rotates clockwise when viewed from the north Galactic pole \citep[e.g.][]{Stanek97, Wegg13}. The presence of a characteristic X-shaped structure of the bulge was first identified by NIR photometry from the COBE satellite \citep{Weiland94}, and later verified looking at the doubling of the red clump \citep[RC, e.g.][]{Nataf10, McWilliam10}. This was then confirmed and shown to become prominent above $\sim 400$ pc from the Galactic plane by \citet{Wegg13}, which mapped the three-dimensional density of the bulge using RC stars. Recent works from \citet{Clarke19} and \citet{Sanders19} confirm the signature of the bar and of the X-shaped structure using star kinematics, deriving absolute proper motions from VVV and {\Gaia}, and creating transverse kinematic maps of the Galactic bulge over a region of $300$ deg$^2$.

These observations are of fundamental importance, since the current morphology of the Galactic bulge can reveal the physical mechanisms responsible for its formation.
Historically,  two main bulge formation mechanisms have been introduced, resulting in the so-called classical bulges and pseudo-bulges \citep{Kormendy04}. Classical bulges are expected to form following the mergers that a galaxy experiences through cosmic time, and they predict an old population of stars. On the other hand, the formation of pseudo-bulges involves dynamical instabilities in the stellar disk that give rise to a barred bulge, which then evolves to an X-shaped structure due to buckling and vertical spreading. This second scenario therefore predicts that the bulge should contain stars with a broader distribution of ages, including younger stars, which would not be expected according to the classical merger scenario. Both types of bulges are observed in nearby galaxies \citep[e.g.][]{Kormendy10}, and they are not mutually exclusive \citep{Erwin15, Erwin21}. Even if the pseudo-bulge scenario seems to be favoured by the observations \citep{Ness13_kinematic, DiMatteo15}, a dependence of the observed kinematics on the metallicity of the stars \citep[e.g.][]{Soto07, Pietrukowicz12, Pietrukowicz15, Portail17, Clarkson18, Kunder20} suggests that the bulge is a composite system. The presence of a classical bulge in the Milky Way is constrained to be very small \citep[e.g.][]{Shen10, Kunder12}, and possibly metal-poor ($[\mathrm{Fe/H}] \lesssim -1$) \citep[e.g.][]{Kirby+13, Arentsen20, Fragkoudi+20}. This points to a more complex evolution scenario than what historically proposed, in which also the influence of the stellar halo and of globular clusters should be included in the global picture \citep[e.g.][]{Sestito+19, DiMatteo19, Massari+19, Horta21}.

Observations of the Galactic bulge are hindered by our location in the Galaxy, but precise absolute proper motions of stars can be used to study the kinematics of the different populations and to isolate bulge members from foreground contaminants rotating coherently in the stellar disk. Using proper motions from the Hubble Space Telescope (HST), \citet{Kuijken02} first separated the foreground and bulge populations in two bulge fields (the low-dust Baade's Window and Sgr I), finding strong evidence for the rotation of the bulge. Other subsequent works used deep HST proper motions with a clean separation of bulge and disk stars \citep[less than $0.2\%$ contamination from the disk,][]{Clarkson08}, providing evidence that at most $3.4\%$ of the bulge sources are younger than 5 Gyr \citep{Clarkson11}. Proper motion cleaning of bulge fields allowed also the first detection of blue straggler stars \citep[BSS,][]{Clarkson11} and of the white dwarf cooling sequence \citep{Calamida14} in the bulge. \citet{Clarkson18} studied the proper motion rotation curves from HST as a function of metallicity for main sequence stars, finding that the metal rich population shows a steeper gradient with Galactic longitude, with a greater rotation amplitude.  Simulations provided by \citet{Gough-Kelly22} predict that young stars ($< 7$ Gyr) rotate more rapidly than old stars near the bulge minor axis, which is consistent with findings by \citet{Clarkson18} in the SWEEPS field.  At larger Galactic longitudes, these populations are expected to trace the underlying density: young stars have orbits aligned with the bar (resulting in a rotation curve with forbidden velocities), while old stars show an axisymmetric velocity distribution \citep{Gough-Kelly22}.

The star formation history of the Milky Way bulge, as constrained by photometry, remains a complicated problem due to the complex foreground populations of the thin and thick disks, variable extinction, spatial depth, and extreme image crowding.  Even after 30 years of HST photometry, definitive constraints on the fraction of intermediate age (defined as $<8$ Gyr old) stars are still not available.  
While the old and red main sequence turn-off (MSTO) observed by \citet{Clarkson08} seems to point to an exclusively old population, as found also by \citet{Zoccali03, Clarkson11, Valenti13, Renzini18, Savino20}, the discovery of a population of young stars seems to challenge these results \citep[e.g.][]{Bensby13, Bensby17}. Looking at high-resolution spectra of microlensed dwarf and subgiant bulge stars, \citet{Bensby17} find that more than one third of the metal rich stars are younger than 8 Gyr. \citet{Hasselquist20}, using a supervised machine learning approach, infer ages for $\sim 6000$ metal-rich bulge stars, finding a non-negligible fraction of stars with ages between $2$ and $5$ Gyr close to the plane of the disk. Recent photometry even claims populations $<1$ Gyr \citep{Saha19}. One possible solution is given by \citet{Haywood16}, which find that, given the range of metallicities observed in the bulge, a uniformly old stellar population would produce a spread in colour at the MSTO, which is not observed in the SWEEPS field. Instead, the presence of stars younger than 10 Gyr in the bulge would explain the more narrow sequence observed. \citet{Barbuy18} later noticed that the simulations of the CMDs in \citet{Haywood16} produce too many bright stars, which are not consistent with the deep HST data. While HST colour-magnitude diagrams are compelling, a notable weakness is in their sampling of tiny pinpoint fields over a narrow range of bulge fields, therefore lacking the survey area and sample size required to detect rare populations that might be present over a wide area. The possibility that the bulge has intermediate age stars is also sustained by the presence of long period, luminous Miras \citep[see e.g.][]{Catchpole16} whose progenitors must logically be present in the form of intermediate age main sequence stars. \citet{Schultheis17}, using APOGEE stars in Baade's Window, find a bimodal distribution, with an  old population at $\sim 10$ Gyr, and a long tail towards younger ages, down to $\sim 2$ Gyr. Looking at proper motion-cleaned CMDs for four HST fields with low reddening, \citet{Bernard18} find that $80\%$ of the stars are older than 8 Gyr, with $10\%$ of the brighter ($V \lesssim 21$) stars being younger than 5 Gyr. A recent work from \citet{Surot19} employs synthetic photometric catalogues to decontaminate the VVV CMDs from foreground disk stars on a field along the bulge minor axis, at $b = -6^\circ$. The authors find that the best fitting model favours a population of stars with ages between $7.5$ Gyr and $11$ Gyr, and that young stars are not present in their observations. Similarly, \citet{Joyce+22} perform a re-determination of the bulge age distribution according to \citet{Bensby17}'s own parameters using MIST isochrones \citep{Choi16, Dotter16}, finding a lower number of stars consistent with being younger than $7$ Gyr, and not finding conclusive evidence for the existence of a bulge population with ages $<5$ Gyr.

The Blanco DECam Bulge Survey (BDBS) is an imaging survey spanning more than $200$ deg$^2$ of the southern Galactic bulge using the Dark Energy Camera at the CTIO-4m telescope, providing photometry calibrated to the SDSS $u$ and Pan-STARSS $grizy$ filters \citep{Rich20, Johnson20}. BDBS photometry reaches a median depth of $i=22.3$ \citep{Johnson20}, or $\sim 3$ mag fainter than the $10$ Gyr old MSTO. Recent works from BDBS include the derivation of a tight colour-metallicity relation for RC stars \citep{Johnson20}, the analysis of the double RC \citep{Lim21}, the investigation of multiple populations in globular clusters \citep{Kader22}, and the derivation of the metallicity distribution function for RC stars \citep{Johnson22}. BDBS optical and near-ultra violet (UV) photometry in the less reddened southern Galactic bulge is a perfect match to {\Gaia} astrometry, whose performance and completeness is highly affected by dust extinction and stellar crowding \citep[e.g.][]{Boubert20}. The most recent data release of {\Gaia} is the early third data release (EDR3), which contains data collected over a period of 34 months \citep{GaiaEDR3}. {\Gaia} EDR3 provides astrometry (positions, parallaxes and absolute proper motions) and photometry (magnitudes in the {\Gaia} $G$ band, and in the blue photometer (BP) and red photometer (RP) bands, $G_\mathrm{BP}$ and $G_\mathrm{RP}$) for $\sim 1.5$ billion stars with $G \lesssim 21$ \citep{Lindegren+21a, Riello21}. Radial velocities are currently available for a subset of $\sim 7$ million bright stars \citep{Katz+19, Seabroke21}.

A complication to the study of the Milky Way bulge is given by the presence of a stubborn population of foreground main sequence stars in the disk whose presence renders age constraints extremely more challenging, as well as high and variable reddening and the complex spatial structure of the bar.  The advent of the {\Gaia} EDR3 catalogue offers a potential breakthrough, enabling use of the proper motion cleaning technique \citep[e.g.][]{Kuijken02, Clarkson08} on a vast scale. We exploit this possibility in this paper, attaching precision EDR3 astrometric measurements to BDBS photometry, with the goal of producing bulge CMDs significantly corrected for foreground disk contamination. 

This paper is organized as follows. In Section \ref{sec:obs} we describe the cross-matching procedure used to identify stars observed by both BDBS and {\Gaia} EDR3, and we show the properties of the resulting catalogue of more than 100 million individual sources. In Section \ref{sec:Baade} we present our proposed method to use {\Gaia} EDR3 astrometry to clean the bulge CMDs, applied to stars within 1 deg from the centre of Baade's window. Section \ref{sec:bulge_fields} then generalizes the method to 127 individual fields in the southern Galactic Bulge, allowing us to map the different stellar populations across an area of $\sim 130$ deg$^2$. In Section \ref{sec:results} we show our results, in terms of observed cleaned bulge CMDs and transverse kinematic maps of the bulge. We also present our results in a wider context, discussing the implications on the age and formation of the Galactic bulge. Finally, in Section \ref{sec:summary}, we summarize our analysis and results.

\begin{figure*}[]
\centering
\includegraphics[width=\textwidth]{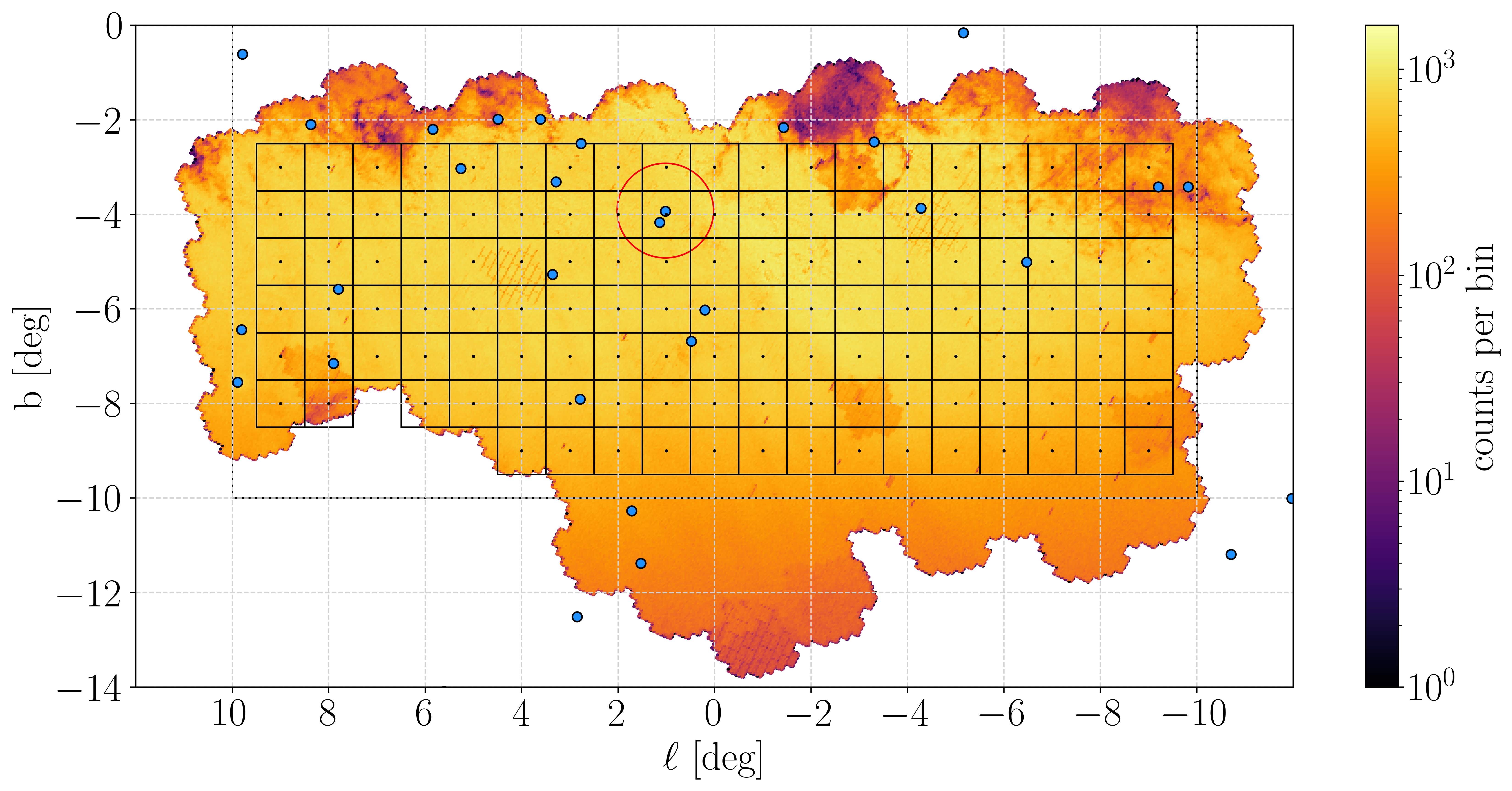}
\caption{Logarithmic density, in Galactic coordinates, of all the sources from the BDBS/{\Gaia} EDR3 matched catalogue. The colour is proportional to the number of stars in each bin. The bins have sizes of $2.26' \times 2.26'$. The dashed box marks the region where we can use the extinction map derived by \citet{Simion17}. The red circle shows the location of the field centred on Baade's window with a radius of $1^\circ$, analysed in Section \ref{sec:Baade}. The black squares show the bulge fields examined in Section \ref{sec:bulge_fields}, each covering an area of $1$ deg$^2$. The blue-filled circles correspond to the known clusters in the field, taken from the catalogue of \citet[][2010 edition]{Harris96}.}
\label{fig:bulge_field}
\end{figure*}

\section{The BDBS / {\Gaia} EDR3 crossmatch} \label{sec:obs}

\begin{figure}[]
\centering
\includegraphics[width=\columnwidth]{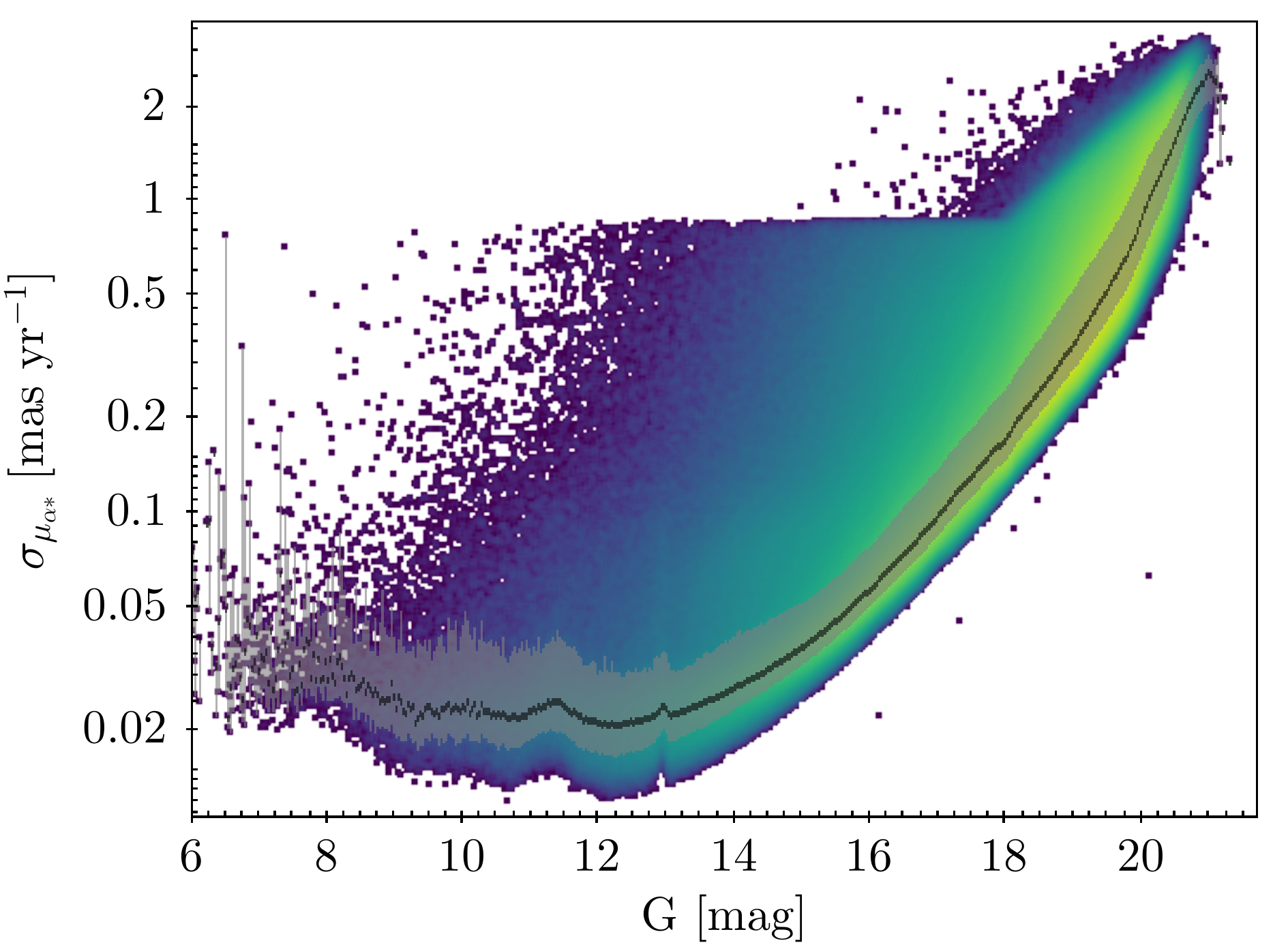}
\caption{Uncertainties in {\Gaia} EDR3 proper motions in right ascension as a function of $G$ magnitude for all the sources in the BDBS / {\Gaia} EDR3 matched catalogue. The median value is shown with a solid black line, and the grey shaded area corresponds to the 1-sigma confidence interval, computed using the 16th and 84th percentiles of the distribution.}
\label{fig:pmra_err_G}
\end{figure}

We crossmatch BDBS to {\Gaia} EDR3, back-propagating {\Gaia} EDR3 coordinates from epoch J2016.0 to J2013.99 (the mean epoch of BDBS observations), using {\Gaia} EDR3 proper motions, when available. We then assign to each BDBS source the nearest {\Gaia} EDR3 star within 1".
This procedure results in a total of $147,496,652$ unique sources in the BDBS/{\Gaia} catalogue. Of these, $100,002,214$ ($\sim 68\%$) have full astrometry from {\Gaia} EDR3\footnote{Applying the same procedure to the second data release of {\Gaia} (DR2) instead of {\Gaia} EDR3 results in $\sim50\%$ fewer matched sources, showing the improvement of {\Gaia} EDR3 in crowded regions.}, and will be the main focus of this paper. The remaining stars have only positions, $G$-band magnitudes, and, in some cases, the colour $G_\mathrm{BP} - G_\mathrm{RP}$. To check the accuracy of the match in position, we estimate $G$ band magnitudes from BDBS $g$ and $i$ photometry, using polynomial transformations comparing {\Gaia} to other known photometric systems\footnote{\url{https://gea.esac.esa.int/archive/documentation/GEDR3/Data_processing/chap_cu5pho/cu5pho_sec_photSystem/cu5pho_ssec_photRelations.html}}. When we compare the estimated and observed {\Gaia} $G$ band magnitudes, we find that $99.7\%$ of the sources have magnitudes consistent within $1$ mag, and $98.9\%$ within $0.5$ mag. By visual inspection of the CMDs, we find that the majority of sources with inconsistent magnitudes have bad photometric measurements. Of the stars with estimated and observed $G$ band magnitudes inconsistent within $1$ mag, we find $\sim 95\%$ of these have $i>18$.  Bulge stars at these fainter magnitudes lie in the region of the CMD which is less sensitive to the {\Gaia} astrometric cleaning, as we show in Section \ref{sec:Baade_astro_clean}, and do not contaminate the bright part of the CMDs above the MSTO.

The distribution in Galactic coordinates of all the sources is shown in Fig. \ref{fig:bulge_field}, where several patterns are evident: low-density regions correspond to dust features (especially at $b \geq -4^\circ$), and to regions of the sky with a lower number of BDBS observations, as is visible for example in the fields at $(\ell, b) = (-3^\circ, -8^\circ)$ and $(\ell, b) = (+4^\circ, -5^\circ)$. The dashed black box corresponds to the $|\ell| \leq 10^\circ$, $b \geq -10^\circ$ region of the sky where it is possible to employ the $1' \times 1'$ reddening map constructed by \citet{Simion17} using RC stars in the VVV survey, which we use in this work to correct the observed magnitudes. The red circle corresponds to the Baade's Window field that is analysed in Section \ref{sec:Baade}, while the black squares correspond to the fields investigated in Section \ref{sec:bulge_fields}. We exclude stars with $b < -2.5^\circ$ because of the higher extinction towards the Galactic plane and due to variations of the reddening on scales smaller than $1'$. The blue dots are the known globular clusters in the area, taken from the list of \citet[][2010 edition]{Harris96}.

Uncertainties in {\Gaia} EDR3 proper motions in the right ascension direction are shown as a function of the observed magnitude in the {\Gaia} $G$ band in Fig. \ref{fig:pmra_err_G} (the trend for the errors in proper motions in the declination direction is equivalent). We see how these values range from $\sim 20$ $\mu$as yr$^{-1}$ for bright sources ($G \lesssim 13$), but then increase steeply with magnitudes, reaching $\sim 1$ mas yr$^{-1}$ at $G = 20$. Uncertainties in {\Gaia} EDR3 parallaxes range from $\sim 20$ $\mu$mas at $G \lesssim 13 $, to $\sim 1$ mas for faint stars with $G = 20$.

\section{Application to stars in Baade's window} \label{sec:Baade}

\begin{figure}[]
\centering
\includegraphics[width=0.8\columnwidth]{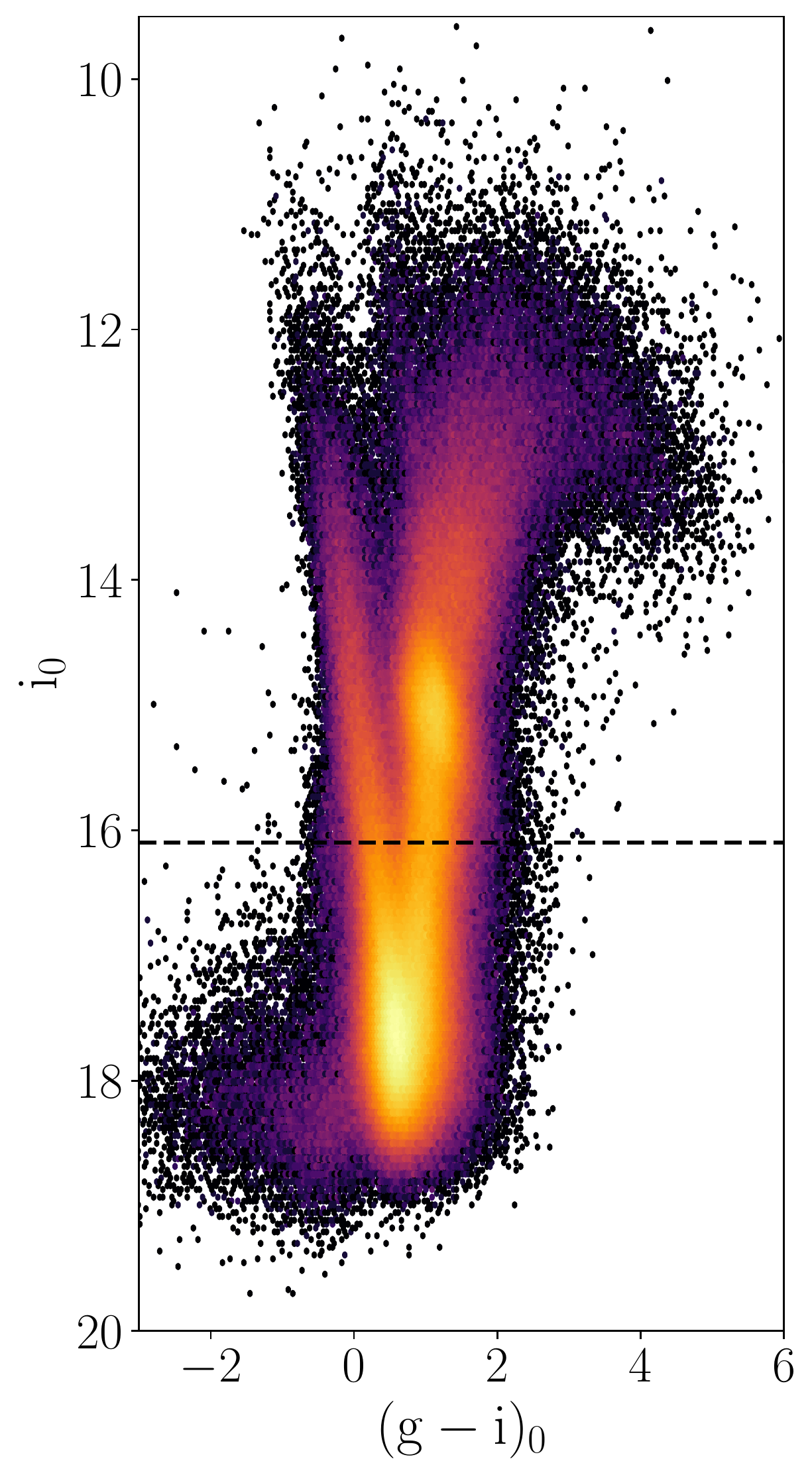}
\caption{Observed dereddened CMD for all the $\sim 1.3 \cdot 10^6$ sources in Baade's window, passing the quality cuts on the {\Gaia} EDR3 astrometry and BDBS photometry. The horizontal dashed line marks the cut at $i_0 = i_{0,\mathrm{max}}-3\sigma_i= 16.1$ which we employ at the end of Section \ref{sec:Baade} to remove the MSTO and compute the ratio of blue to red stars in the field.} 
\label{fig:Baade_CMD}
\end{figure}

\begin{figure*}[]
\centering
\includegraphics[width=\textwidth]{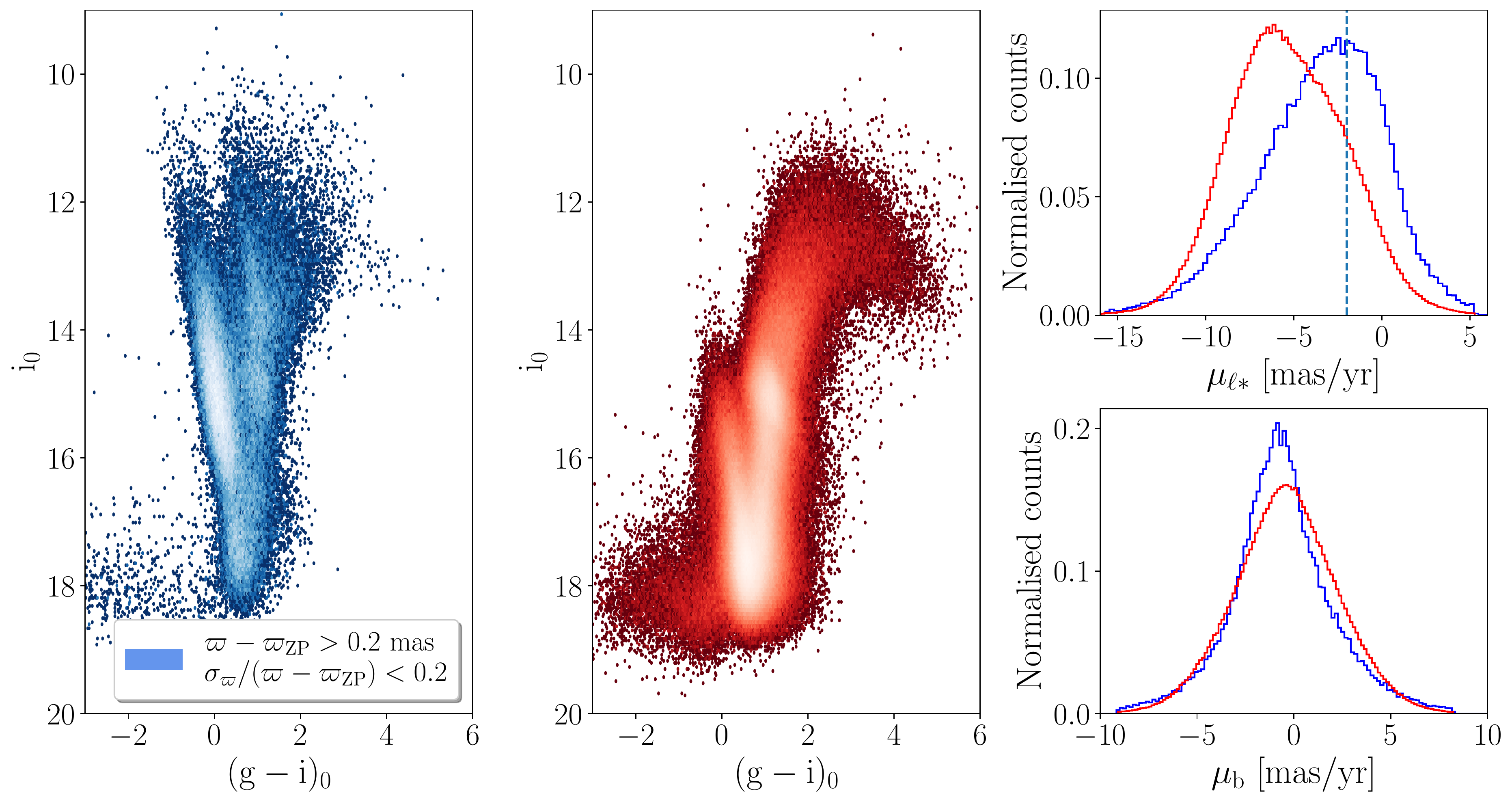}
\caption{Astrometric cleaning procedure using {\Gaia} parallaxes. \emph{Left:} dereddened BDBS CMD for all the $\varpi$-foreground stars in Baade's window, defined as the sources with $(\varpi - \varpi_\mathrm{ZP}) > 0.2$ mas and $\sigma_\varpi/(\varpi - \varpi_\mathrm{ZP}) < 0.2$. \emph{Middle:} BDBS CMD for the $\varpi$-background sources. \emph{Right:} distribution of proper motions in Galactic longitude (top) and latitude (bottom) for $\varpi$-foreground (blue) and $\varpi$-background (red) stars. The vertical dashed line corresponds to the cut in $\mu_{\ell *}$ employed by \citet{Clarkson08} to select bulge members.}
\label{fig:Baade_CMDs_pms}
\end{figure*}

In this section, we focus only on stars in Baade's Window, to describe the procedure we introduce to clean the observed CMDs based on {\Gaia} EDR3 parallaxes and proper motions. Baade's Window is one of the most observed and analysed bulge fields, because of its (relatively) low and uniform interstellar absorption \citep[e.g.][]{Holtzman98}. There are $3,414,177$ sources in the BDBS/{\Gaia} EDR3 catalogue within a radius of $1^\circ$ from the centre of Baade's Window, at $(\ell, b) = (+1.02^\circ, -3.92^\circ)$. This field is shown in Fig. \ref{fig:bulge_field} as a red circle. $2,039,126$ of these stars have full astrometric solution from {\Gaia} EDR3\footnote{In this work, we do not differentiate between sources with $5$-parameter and $6$-parameter solutions from {\Gaia} EDR3 \citep{Lindegren+21a} and we use the astrometric pseudo-colour, when available, only to estimate the parallax zero-point. With full astrometric solution, we refer to the complete determination of the 5 astrometric parameters: coordinates, parallax, and proper motions.}. We remove stars within $5'$ from the centres of NGC 6522 and NGC 6528, the two known globular clusters in the field, and remain with a total of $2,002,241$ sources with full astrometry from {\Gaia} EDR3. These stars will be the main focus of this section. 

\subsection{Astrometric cleaning procedure}
\label{sec:Baade_astro_clean}

We correct {\Gaia} EDR3 parallaxes by subtracting the estimated parallax zero-point $\varpi_\mathrm{ZP}$, following the approach described by \citet{Lindegren+21}. We apply the correction only to sources with {\Gaia} $G$-band magnitudes $6 < G < 21$, with $1.1 \ \mu$m$^{-1}$ $< $ \textsc{nu\_eff\_used\_in\_astrometry} $<1.9$ $\mu$m$^{-1}$ (for the sources with a 5-parameters solution from {\Gaia} EDR3), and with $1.24 \ \mu$m$^{-1}$ $< $ \textsc{pseudocolour} $< 1.72$ $\mu$m$^{-1}$ (for the sources with a 6-parameters solution). For the sources outside these ranges, we assume $\varpi_\mathrm{ZP}=0$. Here, \textsc{nu\_eff\_used\_in\_astrometry} and \textsc{pseudocolour} are the effective wavenumbers used for the astrometric solution when {\Gaia} DR2 photometry in the BP and RP bands was or was not available, respectively \citep{Lindegren+21a}. {\Gaia} DR2 colours were used to calibrate the point spread function for the {\Gaia} EDR3 astrometric solution. In case these were not available, a likely scenario for stars in the densest bulge fields, the astrometric pseudo-colour was estimated using the chromatic displacement of the image centroids. For the sample of stars in Baade's Window, we find $\varpi_\mathrm{ZP} \in [-0.08, 0.02]$ mas, with a median value of $-0.025$ mas. We remind the reader that the parallax zero-point should be subtracted from the observed parallaxes, therefore a mean negative value implies that, on average, stars are closer than the distance implied by the nominal value of their {\Gaia} EDR3 parallax.

We then restrict our sample to the sources with reliable astrometric measurements from {\Gaia} EDR3 and good photometry from BDBS in the $g$ and $i$ bands.  The photometry was cleaned by selecting only stars that have median \textit{chi} and \textit{sharp} parameters residing within one standard deviation of the mean values in the original images.  Furthermore, we restricted the sample to only include stars that did not have unusually high sky values relative to the local background on each image and were detected at least two times in each band.  Further information about the quality flags is provided in \citet{Johnson20}.  We also applied the following selection cuts on the {\Gaia} EDR3 data:
\begin{equation}
    \label{eq:cut_1}
    \sigma_{\mu_{\alpha *}} < 1.5 \ \mathrm{mas} \ \mathrm{yr}^{-1},
\end{equation}
\begin{equation}
    \label{eq:cut_2}
    \sigma_{\mu_{\delta}} < 1.5 \ \mathrm{mas} \ \mathrm{yr}^{-1},
\end{equation}
\begin{equation}
    \label{eq:cut_3}
    |\mu_{\ell*} - \langle\mu_{\ell*}\rangle| < 3 \Delta_{\ell},
\end{equation}
\begin{equation}
    \label{eq:cut_4}
    |\mu_b - \langle\mu_b\rangle| < 3 \Delta_b,
\end{equation}
\begin{equation}
    \label{eq:cut_5}
    \textsc{RUWE} < 1.4,
\end{equation}
where $\sigma_{\mu_{\alpha*}}$ and $\sigma_{\mu_\delta}$ are, respectively, the uncertainties on the proper motions in right ascension and declination, $\Delta_\ell$ and $\Delta_b$ are the means of the uncertainties in the distributions of proper motion in Galactic longitude and latitude in the field, respectively, computed from the 16 and the 84th percentiles of the distributions, and \textsc{RUWE} is the {\Gaia} EDR3 renormalized unit weight error \citep[see][]{Lindegren+21a}. The RUWE is the square-root of the reduced chi-square of the astrometric fit, and a large value is an indication of a bad determination of the astrometry of the star. Possible causes for a large value of RUWE include binarity, variability, and instrumental problems \citep{Belokurov20}. 
A subset of $1,289,972$ stars ($\sim 64\%$) satisfies the criteria listed in Equations \eqref{eq:cut_1} to \eqref{eq:cut_5} and the cuts in BDBS photometry. The dereddened BDBS CMD for these sources, created using the reddening map by \citet{Simion17}, is shown in Fig. \ref{fig:Baade_CMD}. We can see a well-defined RC at $i_0 \sim 15$, the MSTO region at $i_0 \sim 17.6$, a blue sequence at $(g-i)_0 \sim 0$ extending up to $i_0 \sim 12$ caused by foreground main sequence stars, and the population of putative 'blue loop' stars at $i_0 \sim 12$, $(g-i)_0 \sim 0.5$ identified by \citet{Saha19} in the same bulge field (see their Figure 15). 

To verify that the astrometric cuts do not introduce a bias in the CMDs, we inspect the CMD comprising the sources which do not satisfy the {\Gaia} cuts in equations \eqref{eq:cut_1} to \eqref{eq:cut_5}. We find that the sources excluded are on average fainter and occupy unexpected regions of the HR diagram, such as the cloud at $(g-i)_0 \sim -1$ and $i_0 \sim 18.5$, and a plume at $(g-i)_0\sim 3$ and $i_0 \sim 15.5$. The rest of the CMD is covered uniformly, allowing us to conclude that the cuts on proper motion do not significantly bias our results.

Even if {\Gaia} EDR3 parallaxes are not precise and accurate enough to provide reliable distances to individual stars in the bulge (and in general beyond a few kpc from the Sun), they can be used to remove obvious foreground stars, which might otherwise contaminate our selection of bulge members. We define as $\varpi$-foreground stars all the sources satisfying:
\begin{equation}
    \label{eq:par_1}
    \varpi - \varpi_\mathrm{ZP} > 0.2 \ \mathrm{mas} 
\end{equation}
and:
\begin{equation}
    \label{eq:par_2}
    \sigma_\varpi / (\varpi - \varpi_\mathrm{ZP}) < 0.2 \ .
\end{equation}
The first condition\footnote{Results do not depend on the exact value of this parameter, since the great majority of stars with precise {\Gaia} parallaxes are within a few kpc from the Sun.} selects all the stars with estimated heliocentric distances within $5$ kpc, while the second selection ensures that we consider only the stars with uncertainties small enough that an accurate distance can be determined by inverting the observed {\Gaia} parallax \citep[see][]{BailerJones15}. For relative errors in parallax above $\sim 20\%$, the inverse of the observed parallax is a biased estimator for the distance of a star, and a Bayesian approach involving the use of prior probabilities on the distance of a star should be implemented \citep[e.g.][]{BailerJones18, Luri18, BailerJones21}.

The observed CMD for the $\sim 7.7 \cdot 10^4$ $\varpi$-foreground stars is shown in the left panel of Fig. \ref{fig:Baade_CMDs_pms}. A well-defined blue sequence is observed centred at $(g-i)_0 \sim 0$, which is composed of nearby disk stars. A population of nearby bright giant stars is also clearly visible at $(g-i)_0 \sim 1$. The middle panel of Fig. \ref{fig:Baade_CMDs_pms} shows instead the observed CMD for the $\varpi$-background sources, defined as the original sample of stars minus the $\varpi$-foreground objects. This sample consists of $\sim 1.2 \cdot 10^6$ stars. We can see how the foreground blue disk population is largely suppressed, but it is still evident at fainter magnitudes $i_0 \gtrsim 14.5$ and $(g-i)_0 \sim 0.5$. These stars are possible foreground contaminants, which are not classified as such because of the large uncertainties in parallax due to their faintness (see equation \ref{eq:par_2}). Nevertheless, we note how almost all of the bright blue sources disappear when applying the cuts in parallax, including the bright ``blue loop" population claimed by \citet{Saha19}, as already shown in \citet{Rich20}. The absence of blue loop stars in the CMDs will be further discussed in Section \ref{sec:Baade_isochrones}. 

\begin{table}
\centering
\caption{Means and covariance matrices of the two bivariate Gaussian distributions, as given by the GMM applied to {\Gaia} EDR3 proper motions of $\varpi$-background stars in Baade's Window (see Fig. \ref{fig:Baade_pms_GMM}).}
\label{tab:Baade_GMM}
\begin{tabular}{c c c}
	\hline\hline
    Parameters & $\mathcal{N}_\mathrm{bulge}(\mu_{\ell *}, \mu_b)$ & $\mathcal{N}_{\mathrm{\mu-foreground}}(\mu_{\ell *}, \mu_b)$ \\
    \hline
    weight & $0.750$ & $0.250$ \\
    $\bar{\mu}_{\ell*}$ [mas yr$^{-1}$] & $-6.525$ & $-1.781$ \\
    $\bar{\mu}_{b}$ [mas yr$^{-1}$] & $-0.235$  & $-0.921$ \\
    $\sigma_{{\mu}_{\ell*}}$ [mas yr$^{-1}$] & $2.587$ & $2.095$ \\
    $\sigma_{{\mu}_b}$ [mas yr$^{-1}$] & $2.675$ & $2.487$ \\
    $\rho({\mu}_{\ell*}, \mu_b)$ & $-0.097$ & $0.014$ \\
    \hline
\end{tabular}
\end{table}

\begin{figure*}[]
\centering
\includegraphics[width=\textwidth]{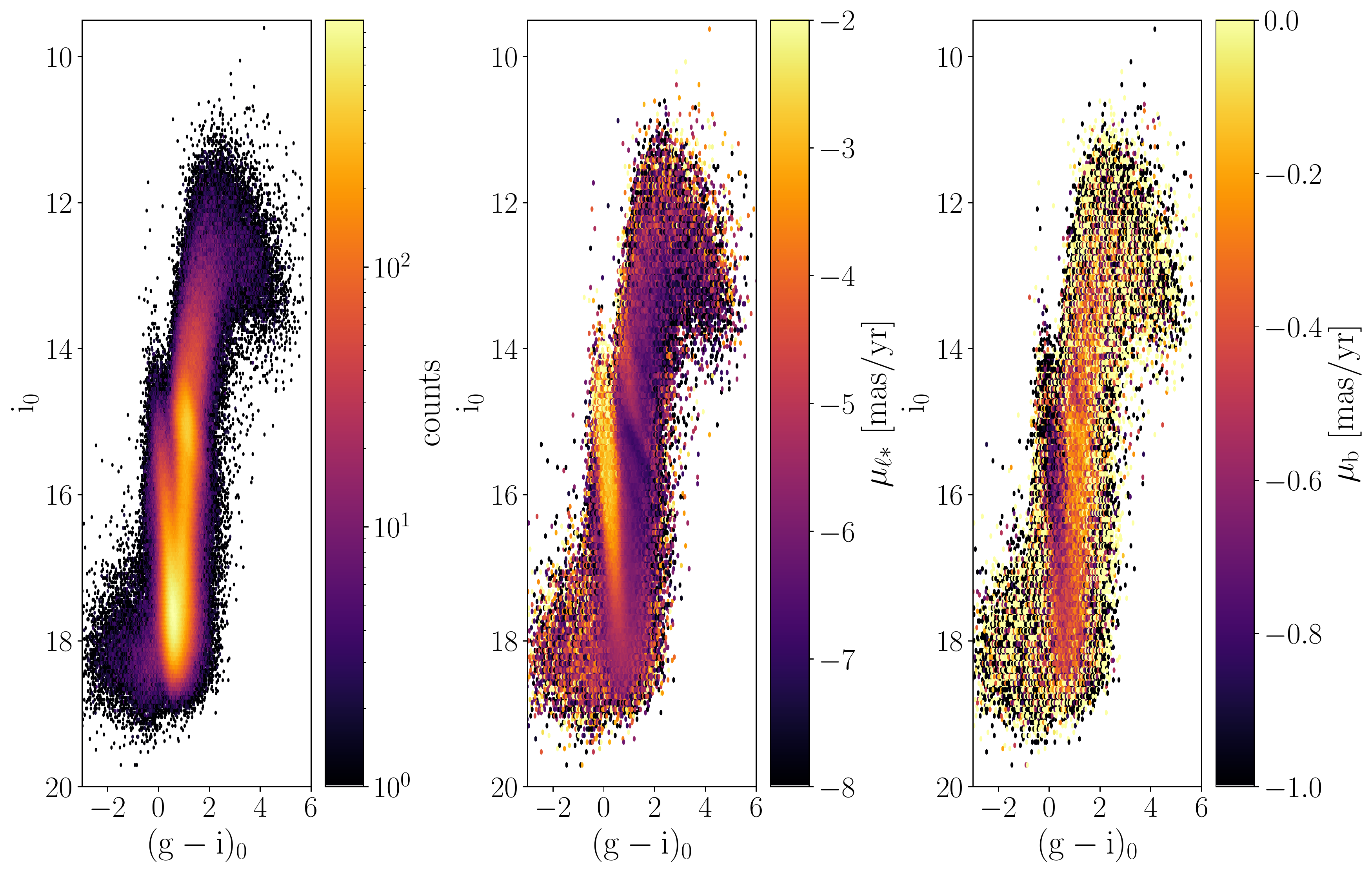}
\caption{\emph{Left:} De-reddened CMD for all the $\varpi$-background sources in Baade's Window, cleaned from foreground contaminants using {\Gaia} EDR3 parallaxes. Colour is proportional to the logarithm of the density of sources. \emph{Middle:} Same CMD, colour-coded by the mean value of the proper motion in Galactic longitude. \emph{Right:} Same CMD, colour-coded by the mean value of the proper motion in Galactic latitude. To compare the proper motion values to their typical uncertainty in {\Gaia} EDR3 (see Fig. \ref{fig:pmra_err_G}), we note that $i_0 = 17$ corresponds roughly to $G = 18.5$.}
\label{fig:Baade_CMDs_bkg_pms}
\end{figure*}

\begin{figure}[]
\centering
\includegraphics[width=\columnwidth]{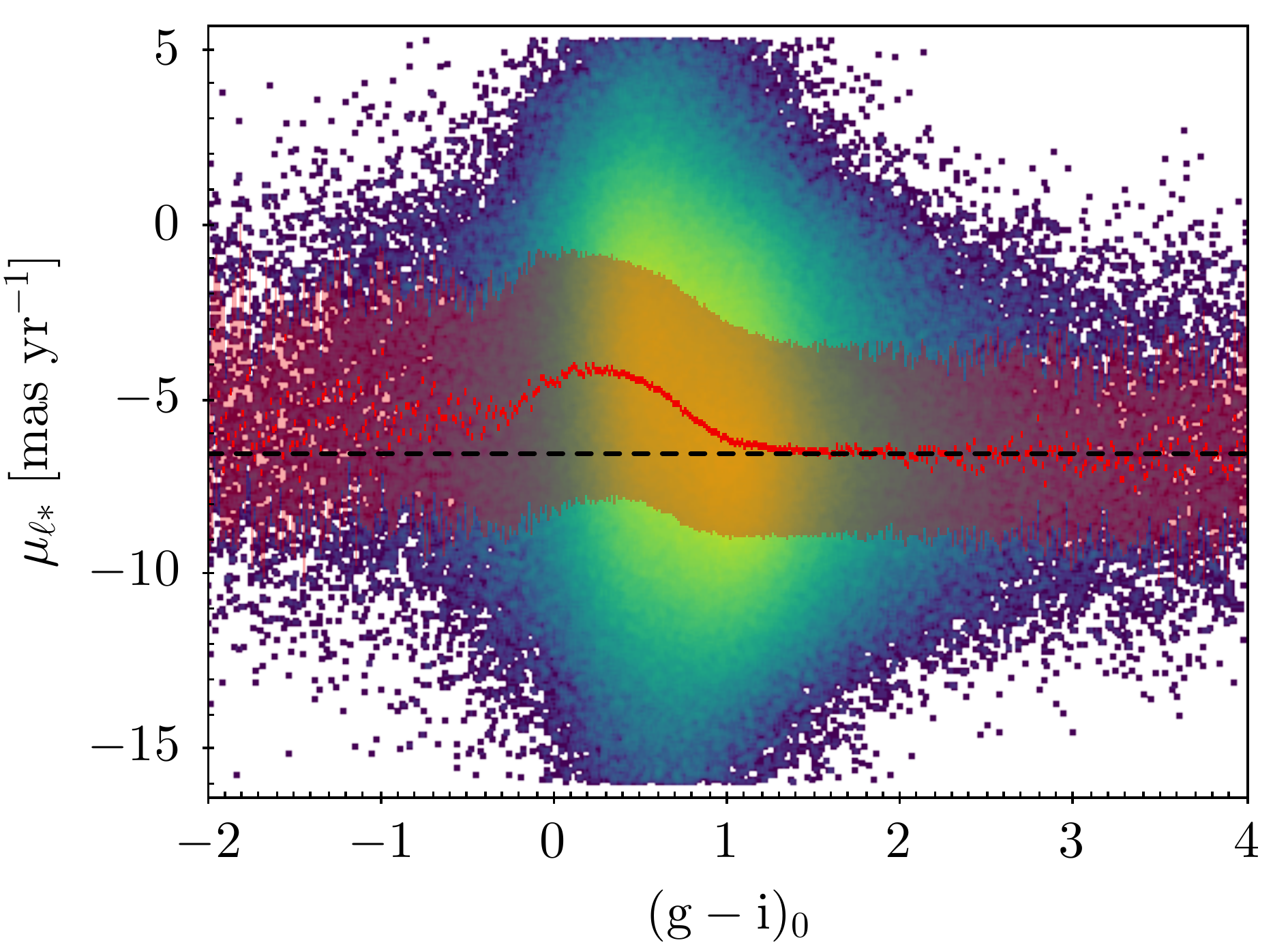}
\caption{Distribution of proper motion in Galactic longitude as a function of colour for the $\varpi$-background sources in Baade's window. The red curve is the median value of $\mu_{\ell *}$ for each bin in $(g-i)_0$, and the width of the red band is computed using the 16th and the 84th percentiles of the distribution of proper motions in each bin. The black dashed horizontal line corresponds to the mean of the Gaussian fitted to the distribution of the bulge stars.}
\label{fig:Baade_gi0_pm}
\end{figure}

\begin{figure}[]
\centering
\includegraphics[width=\columnwidth]{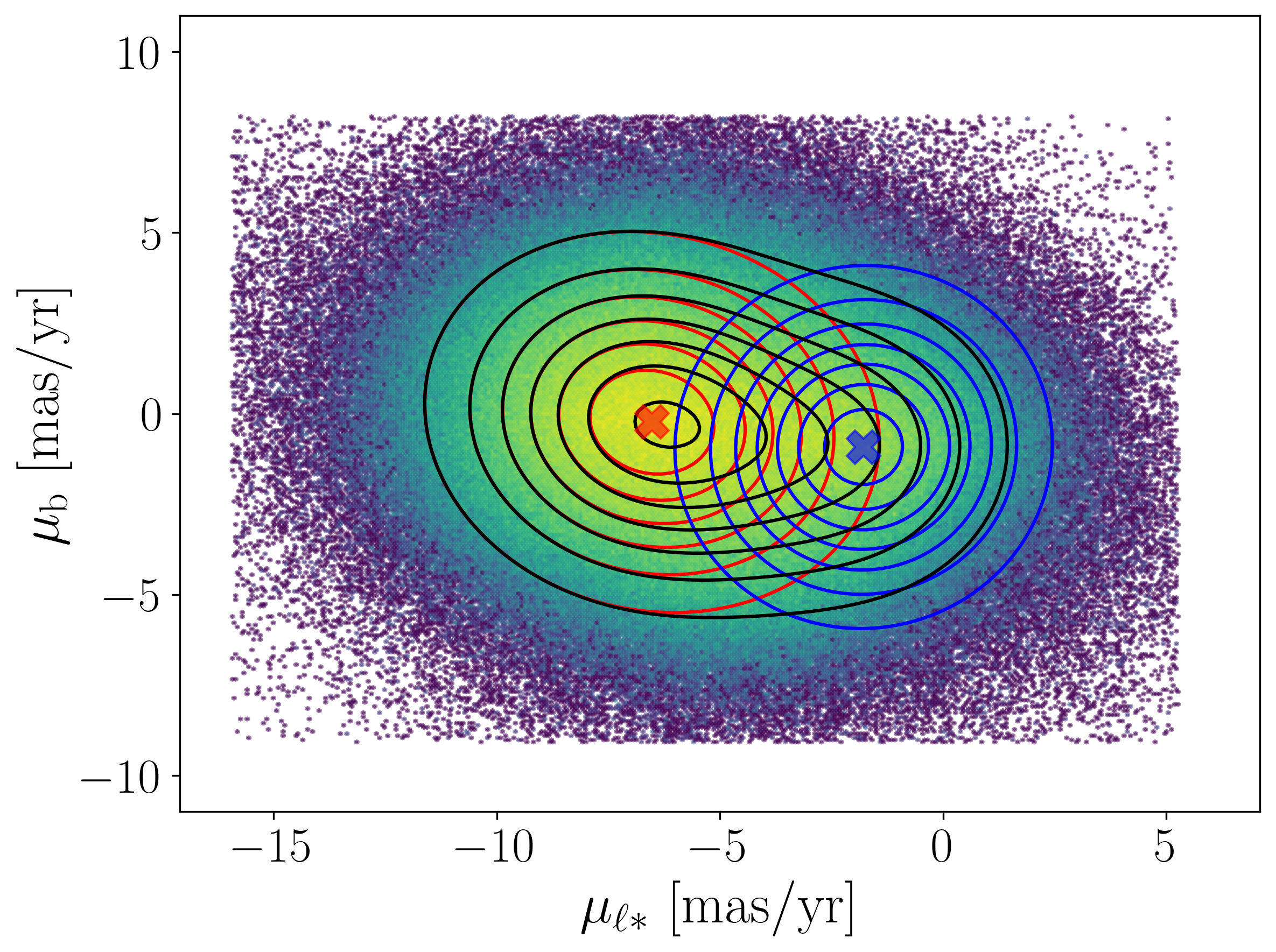}
\caption{Result of the GMM applied to the distribution of Galactic proper motions for $\varpi$-background stars in Baade's window. The red and blue bivariate Gaussian distributions correspond to the bulge and to the $\mu$-foreground population, respectively. The black curve is the sum of the two Gaussian distributions, and the crosses mark the position of the mean of the distributions. The parameters of the two distributions are summarized in Table \ref{tab:Baade_GMM}.}
\label{fig:Baade_pms_GMM}
\end{figure}

\begin{figure}[]
\centering
\includegraphics[width=0.7\columnwidth]{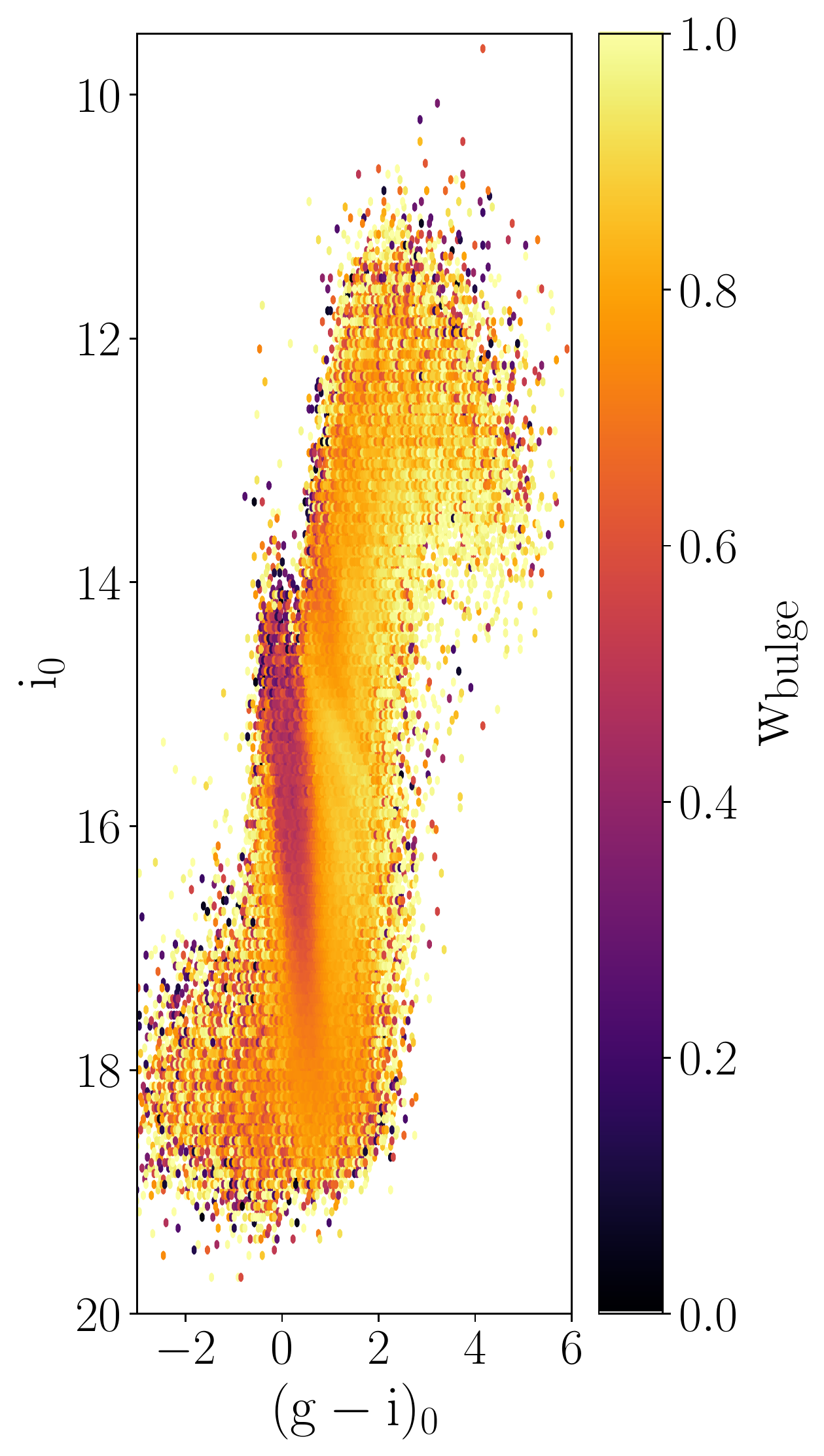}
\caption{De-reddened CMD for all the $\varpi$-background sources in Baade's Window, colour-coded by the probability of bulge membership according to {\Gaia} EDR3 proper motions (eq. \ref{eq:weights}).}
\label{fig:Baade_CMD_pbulge}
\end{figure}

\begin{figure*}[]
\centering
\includegraphics[width=\textwidth]{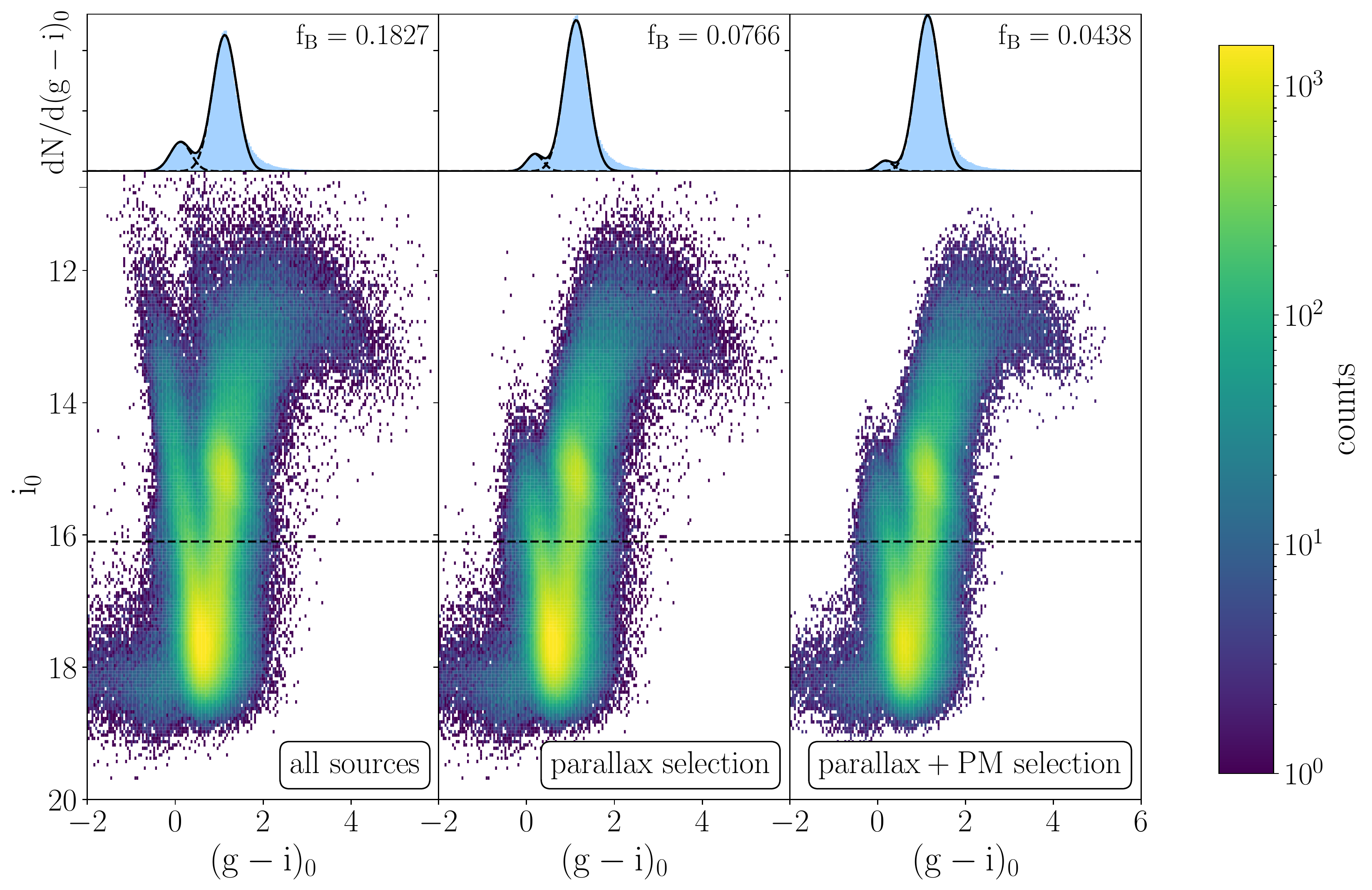}
\caption{Density plots illustrating the parallax and proper motion cleaning of the dereddend CMD for stars in Baade's window. \emph{Left:} all sources with reliable BDBS photometry and {\Gaia} EDR3 astrometry. \emph{Middle:} same as middle panel in Fig. \ref{fig:Baade_CMDs_pms}, $\varpi$-background sources surviving the {\Gaia} EDR3 parallax cuts. \emph{Right:} subset of the $\sim 3.3 \cdot 10^5$ sources with {\Gaia} EDR3 proper motions consistent with a bulge membership, according to the GMM. The blue histograms above the CMDs represent the corresponding normalized distributions of $(g - i)_0$ for stars brighter than $i_{0,max}-3\sigma_i = 16.1$ (shown as a horizontal dashed line in all three plots). The dashed curves correspond to the two fitted Gaussian distributions, and the black curve to their sum. The corresponding value of $f_\mathrm{B}$ (the ratio between blue and red stars) is reported in each panel.}
\label{fig:Baade_CMDs}
\end{figure*}

Given the presence of a faint foreground population which is not easily removed using {\Gaia} EDR3 parallaxes only, we rely on the more precise {\Gaia} EDR3 proper motions, which allow us to further clean the observed CMD to a larger volume. Proper motions in Galactic coordinates $\mu_{\ell*}$ and $\mu_b$ are shown in the right-most panels of Fig. \ref{fig:Baade_CMDs_pms}, in blue and red for the $\varpi$-foreground and $\varpi$-background sources, respectively. As already discussed in e.g. \citet{Kuijken02, Clarkson08, Calamida14, Bernard18, Terry+20}, $\mu_{\ell*}$ can be used to distinguish efficiently between foreground stars and bulge stars, even if there is a clear overlap between the distributions, at $\mu_{\ell*} \sim -5$ mas/yr. A bump in the red distribution is also evident at $\mu_{\ell *} \sim -2$ mas/yr, due to the contamination by foreground objects. The dashed vertical line at $\mu_{\ell *} = -2$ mas/yr corresponds to the threshold value used by \citet{Clarkson08} to isolate bulge members. The distributions in proper motions along Galactic latitude $\mu_b$ are instead more similar, even if for $\varpi$-foreground stars it peaks towards slightly lower (more negative) values. The power of proper motions to identify bulge stars is further shown in Fig. \ref{fig:Baade_CMDs_bkg_pms}, where we present the CMDs for all the $\varpi$-background sources (left panel), colour-coded by the mean value of the proper motion in Galactic longitude and latitude (middle and right panels, respectively). From these plots, it is clear that foreground stars belonging to the blue disk sequence have on average higher values of $\mu_{\ell *}$ and lower values of $\mu_b$. Fig. \ref{fig:Baade_gi0_pm} further shows the power of the proper motion in Galactic longitude to distinguish between blue and red stars. Here, we plot $\mu_{\ell *}$ as a function of the extinction-corrected colour $(g-i)_0$. The red line corresponds to the median value of proper motion for each colour bin, and the red shaded region is computed from the 16th and 84th percentiles of the distribution in each bin. We observe a clear increase in the median value of $\mu_{\ell *}$ for $(g-i)_0 \sim 0.5$, corresponding to blue stars. The value for redder stars then reaches a plateau for $\mu_{\ell *} \sim -6.5$ mas yr$^{-1}$.

Instead of using a single value of $\mu_{\ell *}$ for distinguishing between the two populations, we fitted the two-dimensional distribution in $(\mu_{\ell *}, \mu_b)$ for the $\varpi$-background stars using a Gaussian Mixture Model (GMM) with two components, using the \textsc{sklearn} implementation \citep{scikit-learn}\footnote{We also experiment using the extreme deconvolution algorithm \citep{Bovy11}, which takes into account the full proper motion covariance matrix for each star, using the \textsc{astroML} implementation \citep{astroML}. We find the results to be indistinguishable to those obtained using the simpler GMM, and therefore we decide to choose the latter, simpler method. This follows from the fact that the observed dispersion in proper motions is dominated by the intrinsic velocity dispersion of the stars, and not by the measurement uncertainties (see Fig. \ref{fig:pmra_err_G}).}. We initialize one Gaussian to the mean proper motion we obtain for giant stars only, defined as background stars with $i_0 \leq 15.5$ (this cut, as shown in the middle panel of Fig. \ref{fig:Baade_CMDs_pms}, excludes the great majority of stars belonging to the blue foreground sequence). The other Gaussian is instead initialized to $(\mu_{\ell *}, \mu_b) = (0, 0)$ mas yr$^{-1}$. The parameters of the resulting best-fitting bivariate Gaussian distributions are presented in Table \ref{tab:Baade_GMM}, and the distributions are shown in Fig. \ref{fig:Baade_pms_GMM}, with a red (blue) curve for bulge (foreground) stars. We define the bulge subset as the subset of stars with proper motions consistent with belonging to the red Gaussian (lower values of $\mu_{\ell*}$), and the $\mu$-foreground sample as the subset of sources with proper motions consistent with belonging to the blue Gaussian. We see that the mean value for the bulge Gaussian, shown with a black horizontal dashed line in Fig. \ref{fig:Baade_gi0_pm}, corresponds to the expected value for giant stars. We note that the proper motion dispersions measured for the bulge and disk components, reported in Table \ref{tab:Baade_GMM}, are similar to those reported in HST studies for this and nearby fields, suggesting that {\Gaia} proper motion uncertainties are not significantly impacting our measured dispersions.

To further clean the CMDs and reduce the contamination from foreground stars, we can then assign to each star a weight equal to its probability of belonging to the bulge, according to the {\Gaia} EDR3 proper motions:
\begin{equation}
    \label{eq:weights}
    w_\mathrm{bulge}(\mu_{\ell *}, \mu_b) = \frac{\mathcal{N}_\mathrm{bulge}(\mu_{\ell *}, \mu_b)}{\mathcal{N}_\mathrm{bulge}(\mu_{\ell *}, \mu_b) + \mathcal{N}_\mathrm{\mu-foreground}(\mu_{\ell *}, \mu_b)} \ .
\end{equation}
Here, we use $\mathcal{N}(x,y)$ to indicate a bivariate Gaussian distribution, evaluated in the point $(x,y)$. Fig. \ref{fig:Baade_CMD_pbulge} shows the observed CMD for all the $\varpi$-background stars, colour-coded by $w_\mathrm{bulge}$. We can clearly see that stars belonging to the blue sequence have lower probability to belong to the bulge given their proper motions (as expected from Fig. \ref{fig:Baade_CMDs_bkg_pms}).

We now use $w_\mathrm{bulge}$ to weight each star by its probability of belonging to the Galactic bulge, according to its {\Gaia} EDR3 proper motions. Fig. \ref{fig:Baade_CMDs} summarizes the step used to clean the CMDs using {\Gaia} EDR3 parallaxes and proper motions. The left panel shows the CMD for all the $\sim 1.3 \cdot 10^6$ sources surviving the {\Gaia} EDR3 and BDBS quality cuts. The middle panel shows the distribution of the $\varpi$-background sources surviving the {\Gaia} EDR3 parallax cuts, while in the right panel we plot the same CMD for bulge stars, where each star is weighted by its probability $w_\mathrm{bulge}$ of belonging kinematically to the Galactic bulge. We find a total of $8.8 \cdot 10^5$ stars to be consistent with belonging to the Galactic bulge, according to their parallax and proper motions. We also note that all the bright blue stars are removed from the cleaned CMD, pointing to the fact that the claim of a strong constituency of very young stars in the bulge is not supported by the results presented here. 
The persistence of a faint blue sequence at $(g - i)_0 \sim 0$ proves that the GMM cleaning is still not perfect, as also evident from Fig. \ref{fig:Baade_pms_GMM} and from the parameters of the Gaussian distributions listed in Table \ref{tab:Baade_GMM}, which show a significant overlap between the proper motions of the two populations. In Appendix \ref{app:cleanest} we show that, when imposing much stricter cuts in proper motions to select bulge stars, $\mu_{\ell *} < -5$ mas yr$^{-1}$, this blue plume becomes less prominent, hinting towards the fact that this might be caused by foreground contaminants.

To quantify the maximum contamination from the foreground population in the clean bulge CMDs, we define the following estimator:
\begin{equation}
    \label{eq:f_cont}
    f_\mathrm{cont} = \frac{{\displaystyle\int} \min\Bigl[\mathcal{N}_\mathrm{bulge}(\mu_{\ell *}, \mu_b), \mathcal{N}_\mathrm{\mu-foreground}(\mu_{\ell *}, \mu_b)\Bigr]d\mu_{\ell *} d\mu_b}{{\displaystyle\int} \mathcal{N}_\mathrm{bulge}(\mu_{\ell *}, \mu_b) d\mu_{\ell *} d\mu_b} \ .
\end{equation}
Here, the numerator corresponds to the overlapping area of the red and blue Gaussian distributions in Galactic proper motions shown in Fig. \ref{fig:Baade_pms_GMM}, while the denominator is the area of the red Gaussian distribution. $f_\mathrm{cont}$ is therefore a measure of the contamination to the clean bulge CMDs from stars with proper motions consistent with belonging to the foreground population. For stars in Baade's window, we find $f_\mathrm{cont} = 0.18$. The contamination from bulge stars to the $\mu$-foreground population is instead higher, $\sim 0.54$, because of the smaller weight of the blue Gaussian with respect to the red one, implying a higher number of bulge stars compared to foreground disk stars in the field.

To further quantify the presence of blue stars in the bulge, we now limit our analysis to stars brighter than the MSTO. We estimate this as the maximum $i_{0, \mathrm{max}}$ of the $i_0$ distribution for $i_0 > 17$ (to remove any possible contribution from the RC). We then fit a Gaussian distribution to this distribution, to estimate its standard deviation $\sigma_i$.  With the further condition $i_0 \leq i_{0, \mathrm{max}} - 3\sigma_i = 16.1$, we are left with a sample of $427,951$ stars with full {\Gaia} EDR3 astrometry. This cut is shown with a horizontal dashed line in Fig. \ref{fig:Baade_CMD} and Fig. \ref{fig:Baade_CMDs}.
We quantify the ratio $f_\mathrm{B}$ between the blue and the red sequence populations in the cleaned CMDs by fitting two Gaussian distributions to the distribution of $(g-i)_0$. This is shown in the normalized histograms of Fig. \ref{fig:Baade_CMDs}, where we see how the local maximum at $(g-i)_0 \sim 0.4$ becomes less prominent with further cleaning, confirming its identification with the disk foreground population. By comparing the areas of the two distributions, we find a fraction of blue to red stars for the cleaned CMD of $f_B \equiv N_\mathrm{blue}/N_\mathrm{red} = 0.0437$. We see how $f_\mathrm{B}$ decreases from $\sim 18\%$ for all the sources, to $\sim 8\%$ when removing the foreground objects using {\Gaia} EDR3 parallaxes (the $\varpi$-background sample), to $\sim 4\%$ when adding the GMM proper motion cleaning (the bulge sample).

\begin{figure}
\centering
\includegraphics[width=0.92\columnwidth]{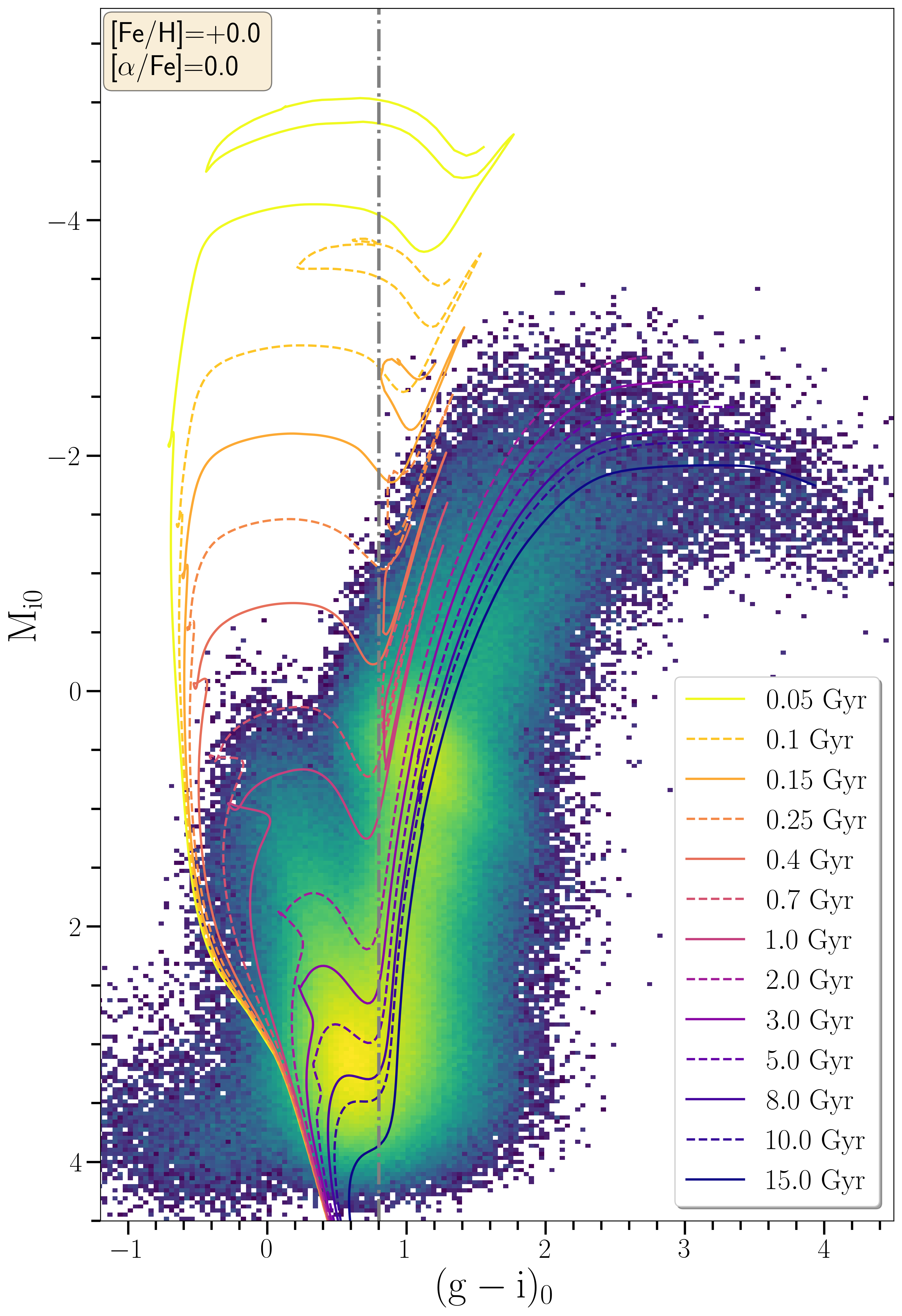}
\caption{\emph{Top}: A selection of MIST isochrones adopting solar composition overlaid on BDBS photometry in Baade's window, after the parallax and proper motion cleaning. The absolute magnitude in the $i$ band $M_{i0}$ is computed assuming a fiducial distance of $8$ kpc for all the stars.
}
\label{fig:isochrones_blueloop}
\end{figure}

\subsection{Location of the Blue Loop according to isochrones}
\label{sec:Baade_isochrones}

As an additional demonstration, the top panel in Figure \ref{fig:isochrones_blueloop} shows a sample of MIST isochrones \citep{Dotter16} overplotted on a clean bulge BDBS CMD in Baade's Window. The models are converted to the PanSTARRS photometry system to match the calibration of BDBS \citep{Johnson20}, assume Solar composition, and span ages from $0.05$ to $15$ Gyr. To compute absolute magnitudes $M_{i0}$ we assume a distance modulus of $14.57$, corresponding to a helio-centric distance of $8$ kpc for all the stars. The chosen value for the distance corresponds to the mean distance of the RC sample in this field, as determined by \citet{Johnson22}, using BDBS data. 

When the theoretical and observed MSTO regions align in the top panel of Fig. \ref{fig:isochrones_blueloop}, it is evident that the stars purported to be blue-loop stars by \citet{Saha19}, observed at $(g - i)_0 \sim 0.5$, $i_0 \sim 12$, are not consistent with the location of the blue-loop feature predicted by the isochrones. The theoretical blue-loop is much brighter than the bulge RGB and falls in a region of the cleaned CMD that does not show any stars. This inconsistency persists in the same direction regardless of the composition assumed in the isochrones (over the range [Fe/H] = -1.8 through +0.5). This morphological difference is hence not explainable by inaccurate metallicity assumptions, nor could it be explained by observational issues that cause systematic offsets. As shown in \citet{Rich20}, these putative blue loop stars are foreground thin disk stars.

It is noteworthy that the purported blue-loop feature disappears entirely as better selection criteria are introduced into the cleaning procedure. Furthermore, any theoretical blue-loop that intersects the observed stars reported in \citet{Saha19} requires a population $<$ 250 Myr in age, and the cleaned CMD strongly rules out such a young main-sequence turn-off. Finally, we remind the reader that the position of the overdensity corresponding to the MSTO is highly affected by {\Gaia} EDR3 incompleteness at the faint end, explaining the mismatch between the theoretical isochrones for ages $\gtrsim 8$ Gyr and the overdensity at $M_{i0} \sim 3.2$ in BDBS data.

\section{Application to different bulge fields}\label{sec:bulge_fields}

\begin{figure*}[]
\centering
\includegraphics[width=0.95\textwidth]{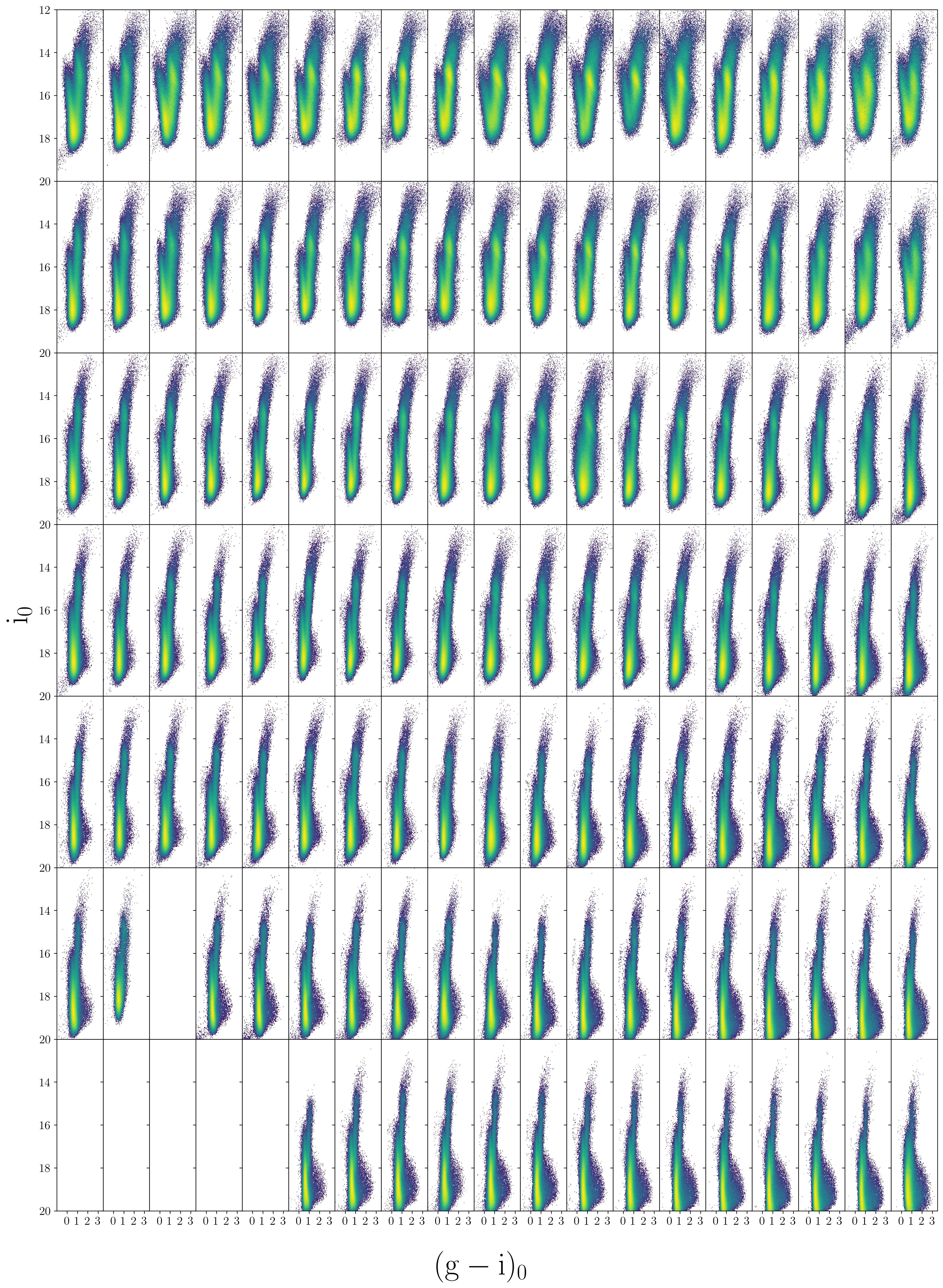}
\caption{Extinction-corrected and foreground-cleaned bulge CMDs for Galactic fields with $|\ell| \leq 9^\circ$ (rows), $-9^\circ \leq b \leq -3^\circ$ (columns), with a step of $1^\circ$. Each field contains all the sources within a square of $1$ deg$^2$. Galactic longitude $\ell$ increases towards the left of the plot, so that the plot in the top left corresponds to $(\ell, b) = (9^\circ, -3^\circ)$, and the one in the bottom right to $(\ell, b) = (-9^\circ, -9^\circ)$, as shown by the black circles in Fig. \ref{fig:bulge_field}.}
\label{fig:CMDs_all_fields}
\end{figure*}

\begin{figure*}[]
\centering
\includegraphics[width=0.95\textwidth]{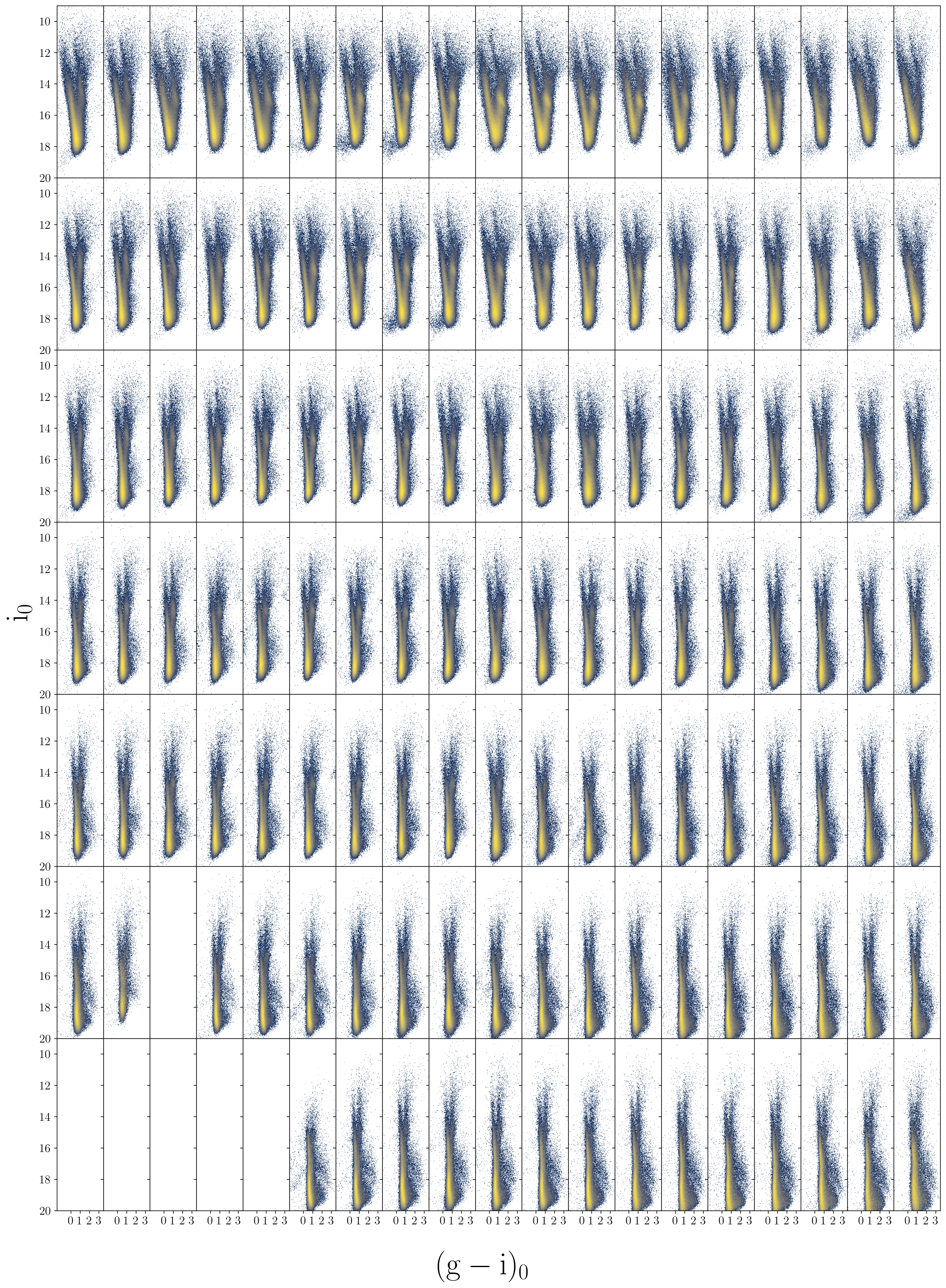}
\caption{Same as Fig. \ref{fig:CMDs_all_fields}, but showing only the foreground sources in each field (selected using both {\Gaia} EDR3 parallaxes and proper motions).}
\label{fig:CMDs_fore_all_fields}
\end{figure*}

\begin{figure}[]
\centering
\includegraphics[width=\columnwidth]{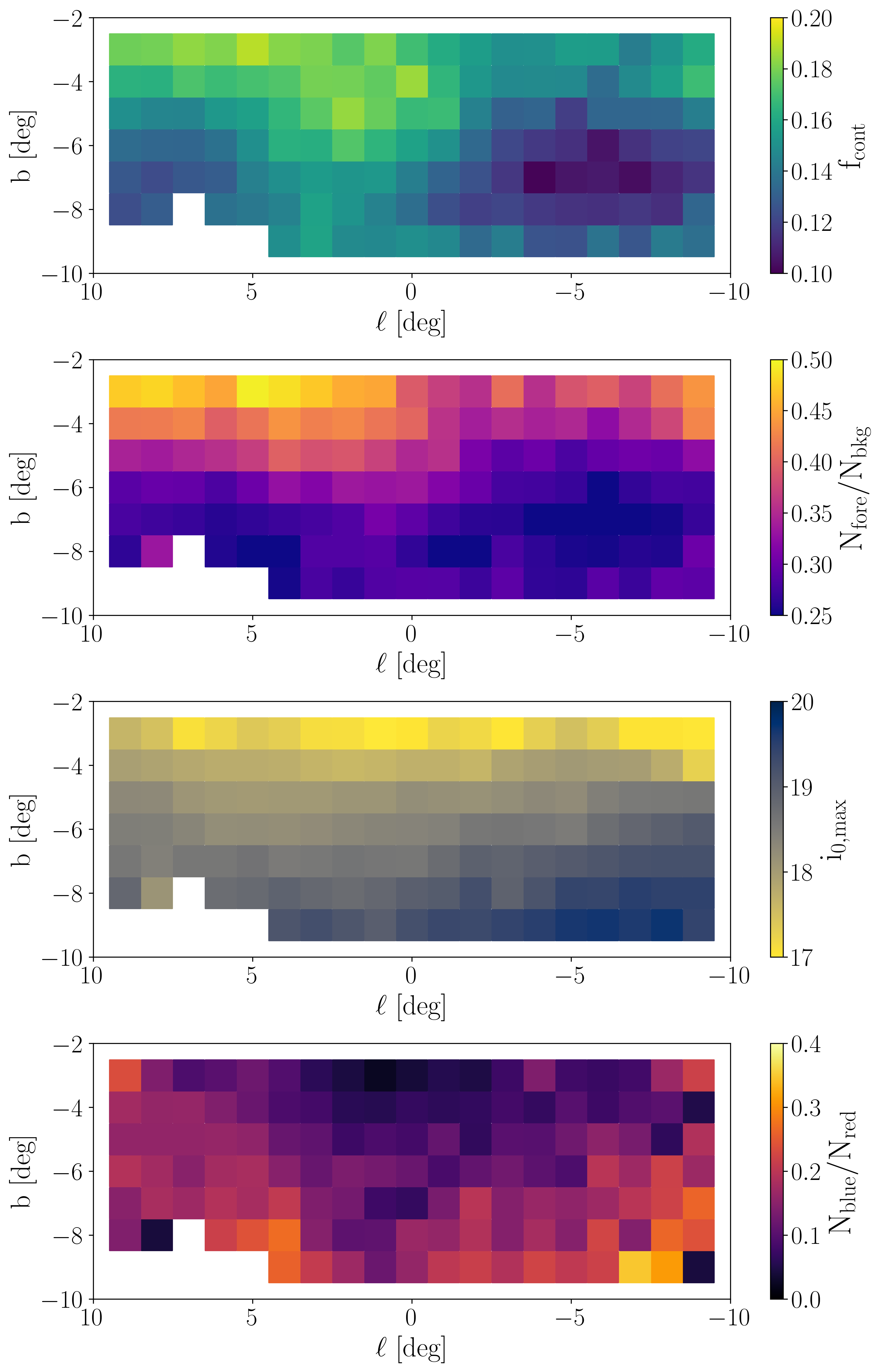}
\caption{\emph{First row:} maximum contamination to the bulge CMDs from stars with proper motions consistent with belonging to the foreground population. \emph{Second row:} ratio of foreground to background stars. \emph{Third row:} Maximum of the $i_0$ distribution (for $i_0 > 17$). \emph{Fourth Row:} Ratio of blue to red stars, when only stars with $i_0 \leq i_{0,\mathrm{max}} - 3\sigma_i$ are considered. All the plots are shown as a function of Galactic coordinates, for the bulge fields shown in Fig. \ref{fig:bulge_field}.}
\label{fig:field_statistics}
\end{figure}

In this section, we apply the methodology introduced and discussed in Section \ref{sec:Baade} to the 127 southern bulge fields shown with black squares in Fig. \ref{fig:bulge_field}. The centres of these fields span a range in Galactic coordinates $|\ell| \leq 9^\circ$, $-9^\circ \leq b \leq -3^\circ$, each covering an area of $1$ deg$^2$. We remove the field at $(\ell, b) = (7^\circ, -8^\circ)$ and those at $\ell \geq 5^\circ$, $b=-9^\circ$ because of the absence (or scarcity) of sources in the BDBS footprint. As mentioned in Section \ref{sec:obs}, the choice of the fields is driven by the range of applicability of the \citet{Simion17} extinction map, which is valid for sources in the VVV footprint, with $|\ell| \leq 10^\circ$, $b \geq -10^\circ$.

For each field, we remove all the sources within $5$ arcmin from the centres of known clusters from \citet[][2010 edition]{Harris96}, and we apply the quality cuts on {\Gaia} EDR3 astrometry and BDBS photometry discussed in Section \ref{sec:Baade}, equations \eqref{eq:cut_1} through \eqref{eq:cut_5}. Then, we remove the $\varpi$-foreground sources, defined as those with precise parallaxes (corrected for the zero-point) consistent with the star being closer than $5$ kpc from the Sun (see equations \ref{eq:par_1} and \ref{eq:par_2}). We then use the GMM to fit the distribution in Galactic proper motions with two bivariate Gaussian distributions, for which we assume that the one with lower (more negative) values of $\mu_{\ell*}$ corresponds to the bulge stars. In Appendix \ref{app:GMM} we explore and discuss the possibility to fit a higher number of Gaussian components to the proper motion distributions. The kinematic weights for each field are then computed using eq. \ref{eq:weights}. Finally, for each field, we estimate the fraction of blue to red stars, fitting Gaussian distributions to the distribution in colour for sources brighter than the MSTO.

\section{Results and discussion} \label{sec:results}

In this Section, we present the results we obtain for the individual bulge fields, following the approach discussed in Section \ref{sec:bulge_fields}.

\subsection{Clean colour-magnitude diagrams}
\label{sec:clean_CMDs}

The clean CMDs for bulge stars in all the fields, obtained weighting their contribution by the corresponding value of $w_\mathrm{bulge}$, are presented in Fig. \ref{fig:CMDs_all_fields}. The plots are presented following the arrangement of the fields shown in Fig. \ref{fig:bulge_field}, so that each column corresponds to a constant value of the Galactic longitude $\ell$ (increasing from right to left), and each row to a constant value of the Galactic latitude $b$ (increasing from bottom to top). The increase in Galactic longitude and latitude in each subsequent field centre is $1^\circ$. Similarly, in Fig. \ref{fig:CMDs_fore_all_fields}, we show the CMDs for all the foreground stars in each field, defined as the combined sample of sources satisfying equations \eqref{eq:par_1} and \eqref{eq:par_2}, and of $\varpi$-background stars weighted by $1 - w_\mathrm{bulge}$ (so the combination of the $\varpi$-foreground and the $\mu$-foreground samples). The difference between Fig. \ref{fig:CMDs_all_fields} and Fig. \ref{fig:CMDs_fore_all_fields} is most evident for the fields closer to the Galactic plane, where the RC is almost absent for the foreground stars, except along the bulge minor axis.

We emphasize again that, because of the significant overlap between the two distributions in proper motions, there is a significant contamination from foreground stars in the bulge fields, as hinted by the presence of a sequence of blue stars. To quantify this possible contamination, we compute $f_\mathrm{cont}$ using eq. \eqref{eq:f_cont} for each bulge field. The result of this is shown in the first panel of Fig. \ref{fig:field_statistics}. We find that $f_\mathrm{cont} \sim 20\%$ at lower latitudes, and decreases to $\sim 10\%$ at further away from the Galactic plane. A clear asymmetry in Galactic longitude, corresponding to the location of the near-side of the bar, is evident at $\ell > 0^\circ$.

In the second panel of Fig. \ref{fig:field_statistics} we show the ratio of the number of foreground to bulge stars in each field. We see how this varies from $\sim 25\%$ for the fields further away from the Galactic plane, and increases to $\sim 50\%$ at lower absolute values of Galactic latitude, where the density of stars in both the stellar disk and the bulge increases. We find this ratio to correlate strongly with the value of $f_\mathrm{cont}$.

In the third panel of Fig. \ref{fig:field_statistics}, we plot $i_{0, \mathrm{max}}$, defined as the maximum of the $i_0$ distribution in each field for $i_0 > 17$, as a function of Galactic coordinates. The $i_0 > 17$ cut is needed to ensure that the maximum of the distribution corresponds to the region of the MSTO, and to remove the possible contamination from the RC, which is particularly prominent in the fields close to the plane, especially at $(\ell, b) = (-3^\circ, -3^\circ)$. The shift in $i_{0,\mathrm{max}}$ is due to two main effects: the presence of dust features, which redden the observed magnitudes and lower the maximum absolute magnitude observable in each field, and the {\Gaia} completeness, which is a strong function of Galactic latitude and crowding of the field \citep[see][]{Boubert20}. Also, the presence of the near-side of the bar induces a dependence on $\ell$, contributing to a decrease in the value of $i_{0, \mathrm{max}}$ at positive values of the Galactic longitude.

We then quantify the ratio of blue to red stars in each bulge field, as previously explained in Section \ref{sec:Baade} for stars in Baade's Window. We select only stars with $i_0 \leq i_{0, \mathrm{max}}-3\sigma_i$ to minimize the contribution from the region around the MSTO, where red giants and nearby disk stars have a similar colour. The bottom panel in Fig. \ref{fig:field_statistics} shows the ratio between the areas of the blue and red fitted Gaussian distributions for the bulge fields explored in this work. We see that these values range from $\sim 2.6\%$ for the fields along the bulge major axis, to $\sim 40\%$ further away from the Galactic plane. We note that the bulge population drops dramatically with increasing distance from the Galactic plane, therefore a lower value of $N_\mathrm{blue} / N_\mathrm{red}$ at low latitudes should not be regarded as an estimate of the contamination from the stellar disk. At lower latitudes, the prominence of the RC (as shown in Fig. \ref{fig:CMDs_all_fields}) results in lower values of $N_\mathrm{blue} / N_\mathrm{red}$, while on the other hand, at higher latitudes, the fraction of red stars is lower, resulting in higher values of $N_\mathrm{blue} / N_\mathrm{red}$. Here we stress that $N_\mathrm{blue} / N_\mathrm{red}$ is computed only for the $\varpi$-background stars (that is, after removing foreground stars using {\Gaia} parallaxes), and therefore the contribution from blue disk stars brighter than the MSTO is further suppressed. This explains the observed anti-correlation between the values of $N_\mathrm{fore} / N_\mathrm{bkg}$ and $N_\mathrm{blue} / N_\mathrm{red}$, shown when comparing the second and fourth panel in Fig. \ref{fig:field_statistics}. A further discussion on the population of blue stars in the fields and their kinematics is provided in Section \ref{sec:blue}.

\subsection{Kinematic Transverse Maps}

\begin{figure}[]
\centering
\includegraphics[width=\columnwidth]{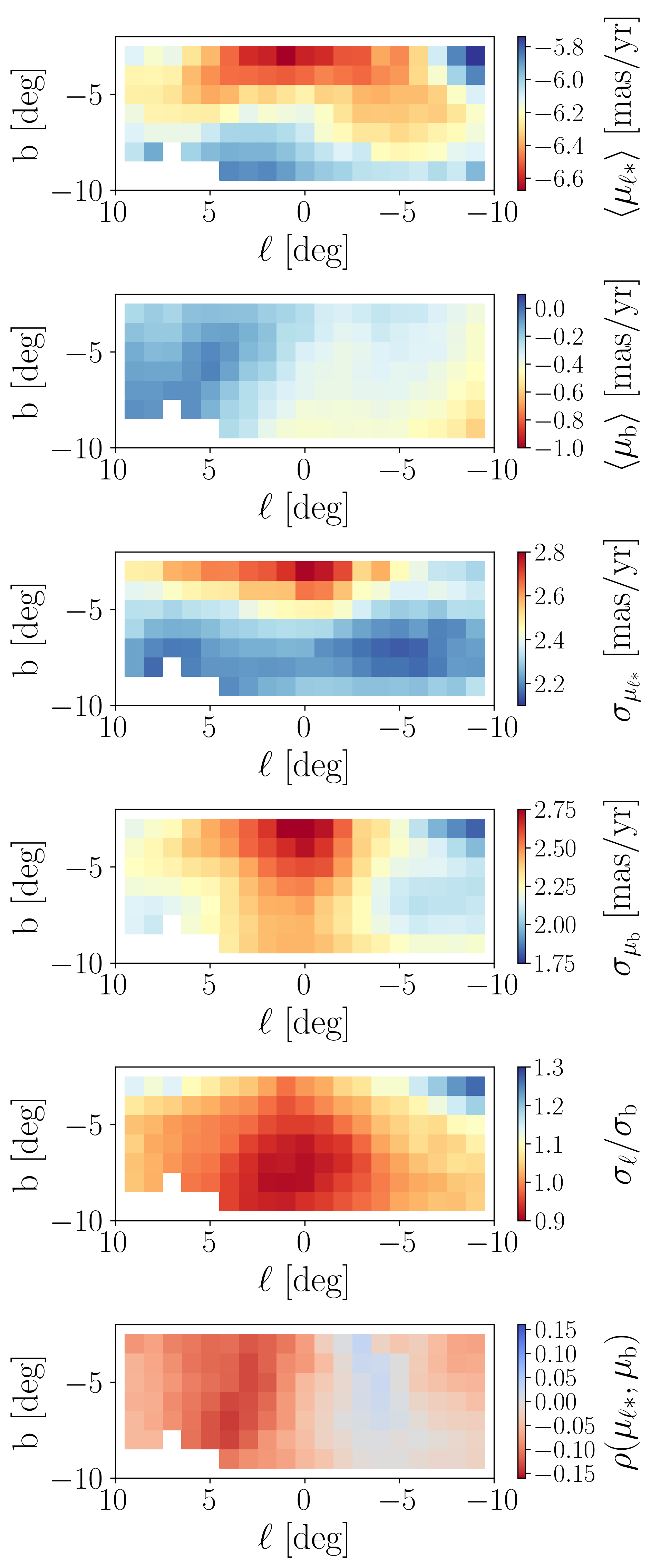}
\caption{Kinematic transverse maps derived from the parameters of the Gaussian distribution of the bulge members fitted to {\Gaia} EDR3 proper motions, for the same bulge fields shown in Fig. \ref{fig:bulge_field}. From top to bottom, we show the mean proper motions in Galactic longitude and latitude, the proper motion dispersions in Galactic longitude and latitude, the dispersion ratio, and the correlation between the proper motions.}
\label{fig:pm_moments}
\end{figure}

We can use the results of the GMM applied to {\Gaia} EDR3 proper motions to investigate the transverse kinematics of stars in the different bulge fields explored in this work. Following the works of \citet{Clarke19, Sanders19} using VVV and {\Gaia} DR2 data, we quantify the bulge kinematics using the mean proper motions of the bulge stars $\langle \mu_{\ell *} \rangle$, $\langle\mu_b \rangle$, the corresponding dispersions $\sigma_{\mu_{\ell *}}, \sigma_{\mu_b}$, the dispersion ratio $\sigma_{\mu_{\ell *}} / \sigma_{\mu_b}$, and the correlation coefficient between the proper motions $\rho(\mu_{\ell *}, \mu_b)$. These quantities are presented in Fig. \ref{fig:pm_moments} for all the bulge fields shown in Fig. \ref{fig:bulge_field}. For an easier comparison, we use the same colours and a similar range as in Figure 10 of \citet{Clarke19}. The first two panels show the projected mean rotation of the bulge stars along $\ell$ and $b$, respectively, and are offset because of the tangential reflex motion of the Sun \citep[see][]{Reid04}. The third and fourth panel, showing the proper motion dispersions, reach a well-defined maximum near the Galactic plane along the minor axis, due to the potential well of the bulge. Also, the observed asymmetries with the Galactic longitude are due to the inclination of the bar, with dispersions being generally smaller (larger) at $\ell < 0$ ($\ell > 0$), when the bar is further away from (closer to) the Sun. The fifth panel shows the dispersion ratio, which is $\sim 1$ along the bulge minor axis, and reaches higher values $\sim 1.2$ near the Galactic plane. The dipole pattern in the correlation between the proper motion components (which becomes a quadrupole pattern when one has access to the Northern Galactic bulge) shows a radial alignment towards the Galactic Centre, and it reaches its maximum amplitudes around $b \sim -6^\circ$, due to the triaxial shape of the bulge.

In general, we find a very good agreement with the results presented in \citet{Clarke19, Sanders19}, both in terms of the VIRAC data \citep{Smith18}, and of the made-to-measure (M2M) model \citep{Portail17, Clarke19}. The magnitude of the dispersions, dispersion ratio, and correlation is also consistent with the study of \citet{Kozlowski+06} in the vicinity of Baade's window. This proves the convergence of the GMM to the expected values, further validating the approach introduced in this paper, and it confirms the kinematic results with the more precise {\Gaia} EDR3 astrometry.

\subsection{Blue stars in the bulge}
\label{sec:blue}

\begin{figure}[]
\centering
\includegraphics[width=\columnwidth]{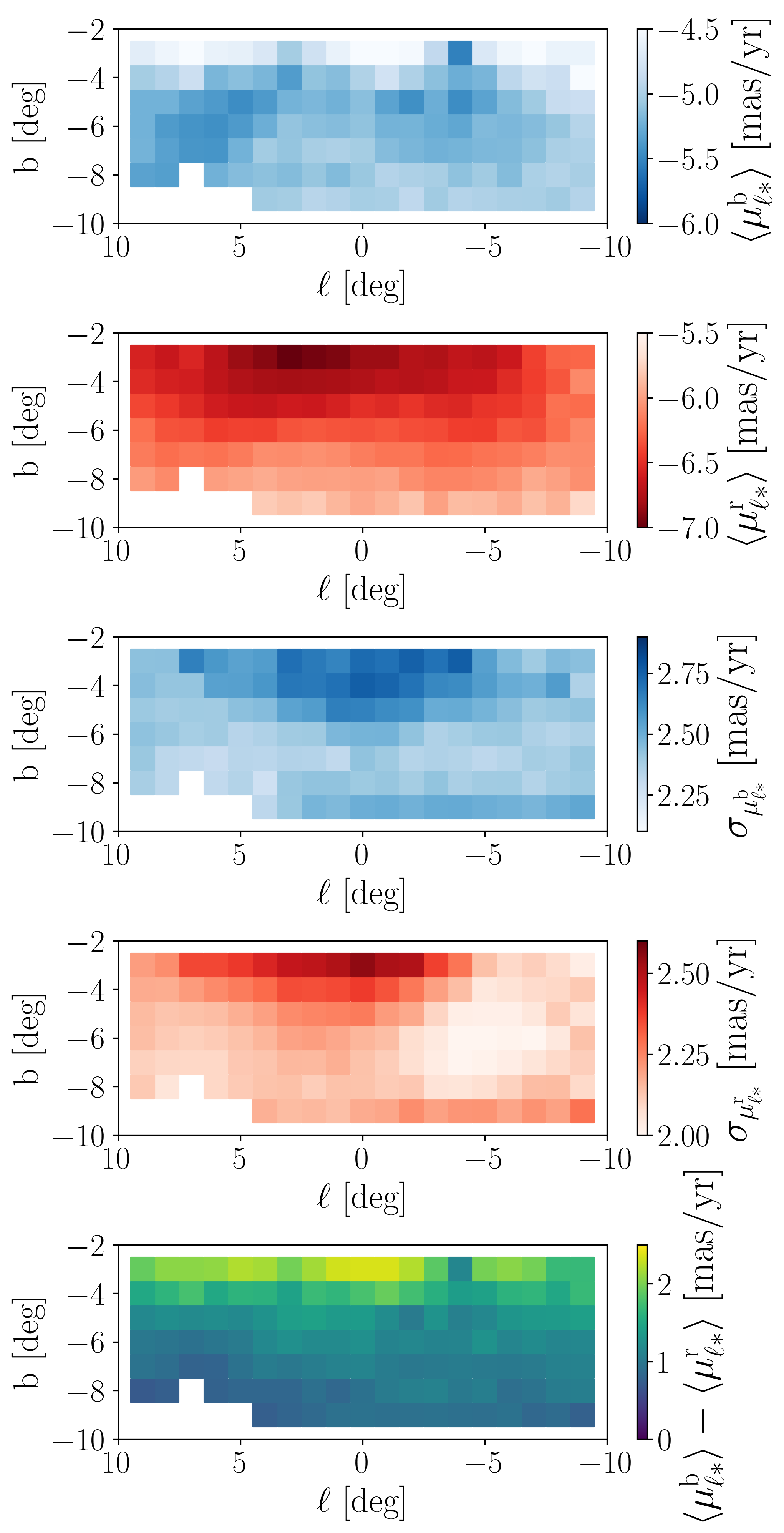}
\caption{From top to bottom, we show the mean proper motion in $\ell$ for blue stars, the mean proper motion in $\ell$ for red stars, the dispersion in proper motions in $\ell$ for blue stars, the dispersion in proper motions in $\ell$ for red stars, and the difference between the mean proper motions in $\ell$ for blue and red stars. All the plots are shown as a function of Galactic coordinates, for the southern bulge fields shown in Fig. \ref{fig:bulge_field}.}
\label{fig:field_statistics_redblue}
\end{figure}

The fourth panel of Fig. \ref{fig:field_statistics} shows that the ratio between blue and red stars in the bulge fields considered in this work is typically around $\sim 3\%$ near the Galactic plane, increasing towards higher absolute Galactic latitudes to $\sim 20\%$. This blue population is clearly visible in the cleaned bulge CMDS shown in Fig. \ref{fig:CMDs_all_fields}, where it is most evident in the fields at $b \geq -4^\circ$ where it clearly separates from the red giant branch.

We now investigate the kinematics of blue and red stars, to understand whether blue stars belong to the bulge population. As explained in Section \ref{sec:clean_CMDs}, we only consider stars with $i_0 \leq i_{0,\mathrm{max}} - 3\sigma_i$. We then fit two Gaussian distributions to the resulting distribution in $(g-i)_0$. Following equation \eqref{eq:weights}, we compute the probability of each star to belong to the blue population as:
\begin{equation}
    \label{eq:P_blue}
    w_\mathrm{blue}\Bigl( (g-i)_0 \Bigr) = \frac{\mathcal{N}_\mathrm{blue}\Bigl( (g-i)_0 \Bigr)}{\mathcal{N}_\mathrm{blue}\Bigl( (g-i)_0 \Bigr) + {\mathcal{N}_\mathrm{red}\Bigl( (g-i)_0 \Bigr)}} \ .
\end{equation}
We can then isolate blue stars by weighting each star in the $\varpi$-background sample by the probability $w_\mathrm{blue}w_\mathrm{bulge}$, and red stars weighting by the probability $(1- w_\mathrm{blue})w_\mathrm{bulge}$.
The first four panels of Fig. \ref{fig:field_statistics_redblue} show the mean proper motions and the proper motion dispersion in Galactic longitude for blue and red stars. The last panel shows instead the difference between the mean proper motion in Galactic longitude between blue and red stars. In these plots we focus on $\mu_{\ell *}$ because it is the proper motion component that is more efficient at separating the bulge and the foreground populations. By comparing Fig. \ref{fig:field_statistics_redblue} to Fig. \ref{fig:field_statistics}, we can see how the population of blue stars shows significantly different trends as a function of Galactic coordinates, with respect to the expected patterns for an X-shaped triaxial bulge \citep{Clarke19, Sanders19}. In particular, we can see how the mean proper motion along Galactic longitude increases with increasing Galactic latitude, which is the opposite of what is predicted for bulge stars. The red stars, on the other hand, show kinematics consistent with what shown in Fig. \ref{fig:field_statistics}.

As a further check, in Appendix \ref{app:cleanest} we show that the faint blue sequence visible in Fig. \ref{fig:CMDs_all_fields}, especially in the fields closer to the Galactic plane, is highly suppressed when imposing the much stricter cut $\mu_{\ell *} < -5$ mas yr$^{-1}$. Together with Fig. \ref{fig:field_statistics_redblue}, this is a strong indication that this peculiar feature is likely to be caused by faint foreground disk stars, whose parallax is too imprecise to have a distance determination from {\Gaia} EDR3, and whose proper motions lay in the overlapping region between the bulge and foreground distributions.

\begin{figure}[]
\centering
\includegraphics[width=\columnwidth]{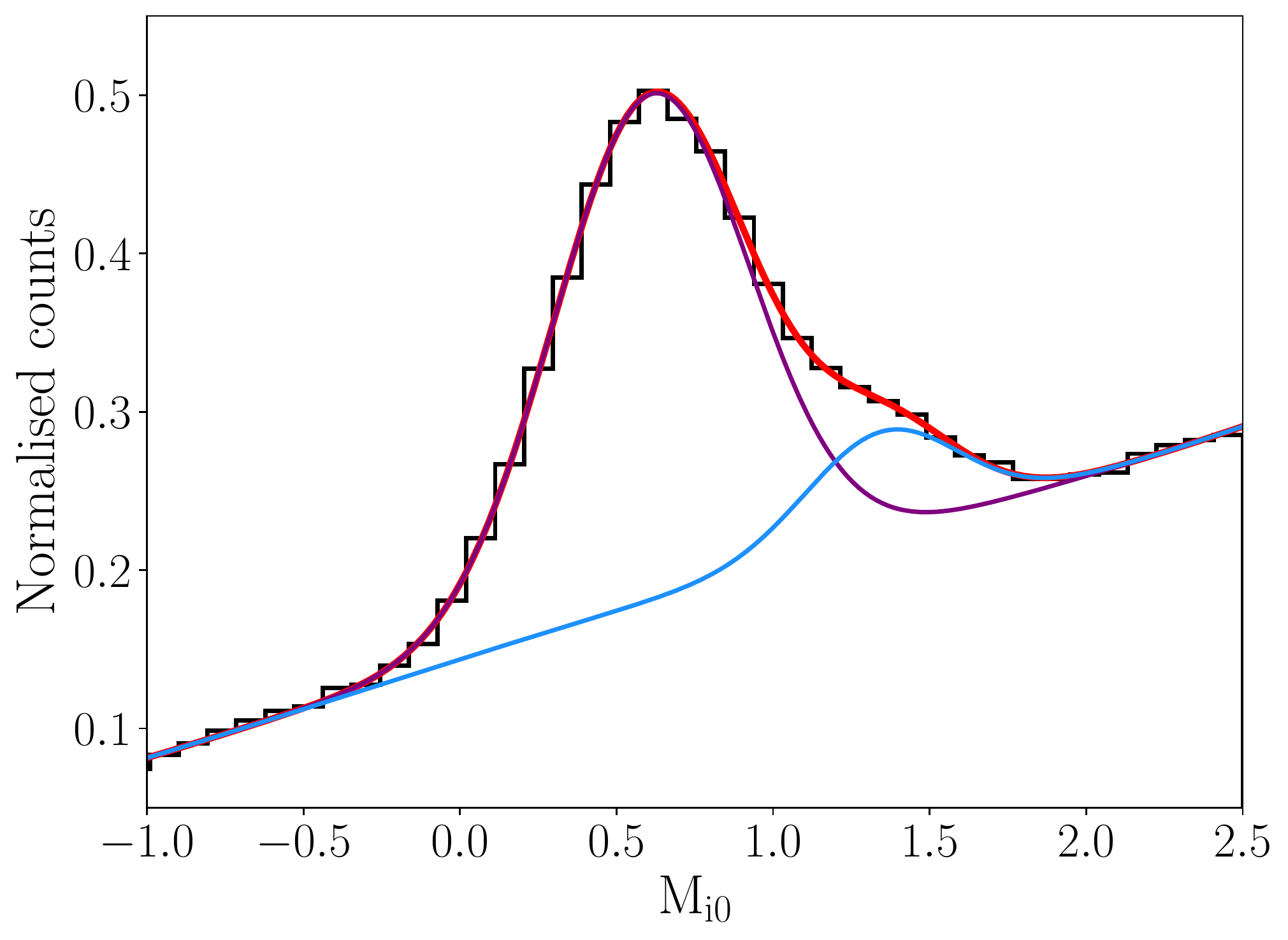}
\caption{Gaussian fits to the absolute magnitude distribution for sources in Baade's window with $(g-i)_0 \geq 0.8$ (dash-dotted vertical line in Fig. \ref{fig:isochrones_blueloop}). The purple curve corresponds to the RC, while the blue one to the RGB bump. Absolute magnitudes are computed assuming a mean distance of $8$ kpc for all the stars.}
\label{fig:Baade_Mi0}
\end{figure}

\begin{figure}[]
\centering
\includegraphics[width=\columnwidth]{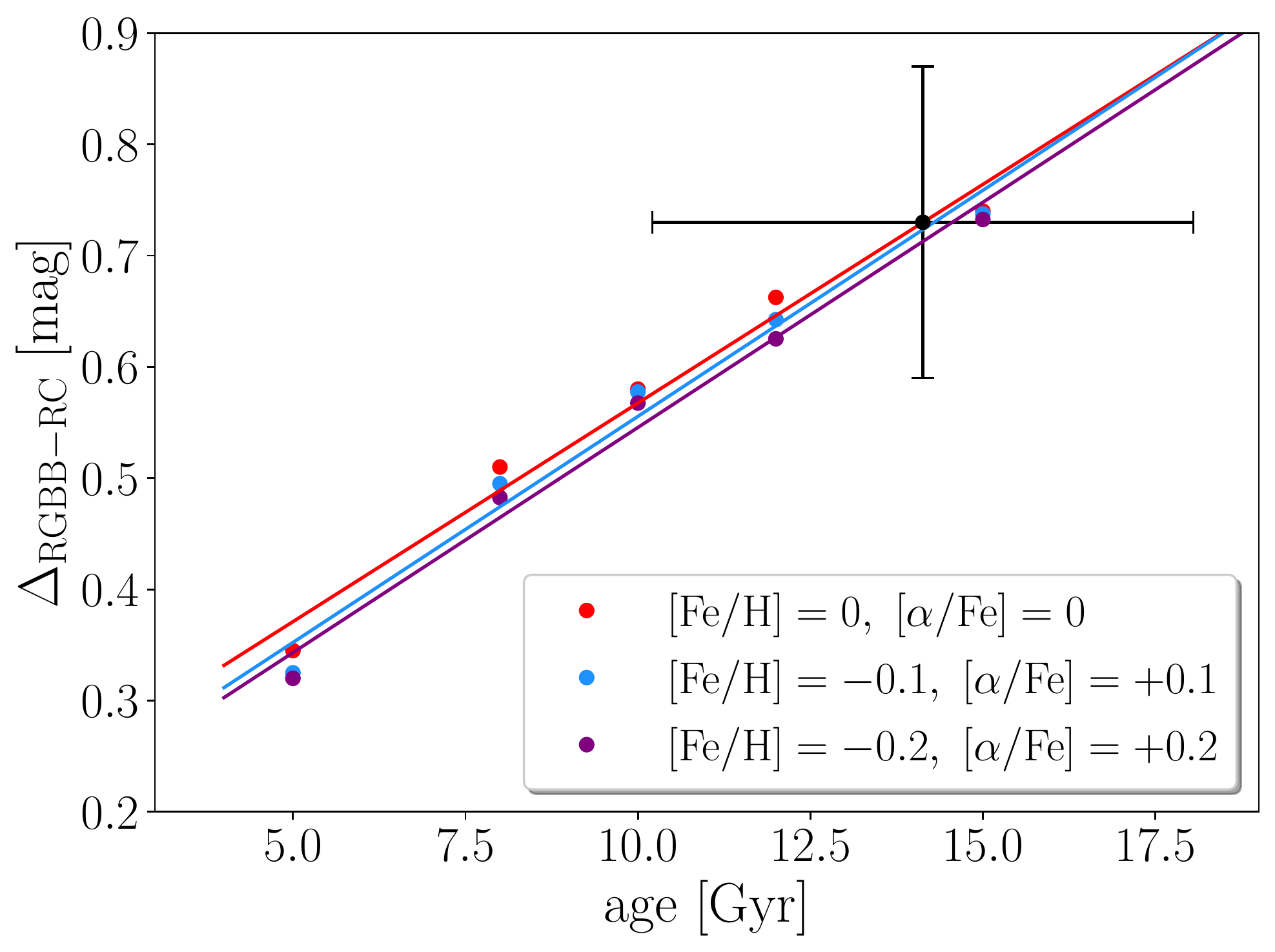}
\caption{Distance between the RGB bump and the RC in the $i$ band as a function of age for MIST isochrones of different metallicities and alpha abundances. The lines are linear fits to the theoretical points. The black dot, with its uncertainties, corresponds to the value obtained in this work for stars in Baade's Window.}
\label{fig:Delta_age_fit}
\end{figure}

\begin{figure}[]
\centering
\includegraphics[width=0.8\columnwidth]{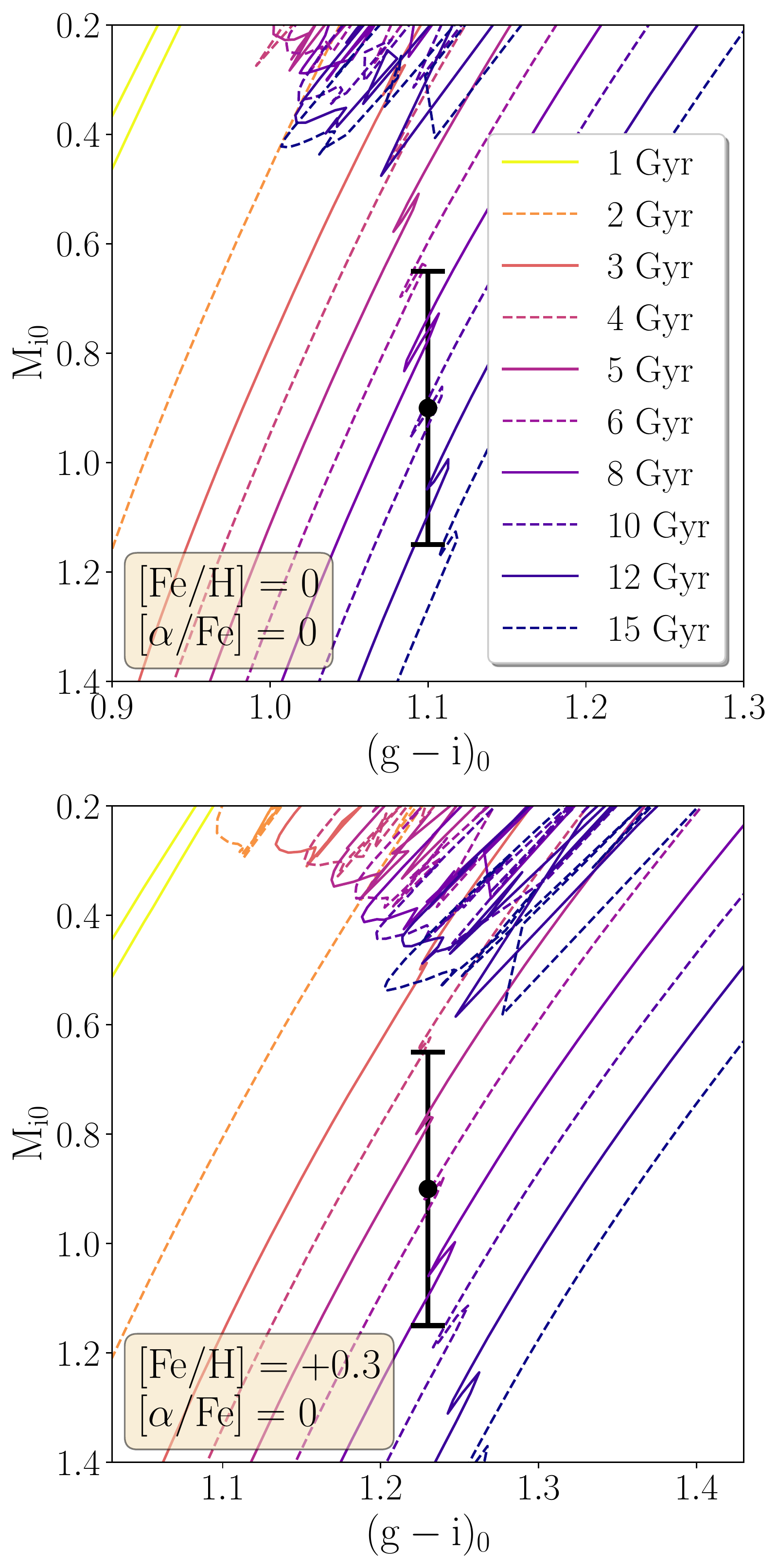}
\caption{Set of MIST isochrones for $\mathrm{[Fe/H]} = 0$ and $\mathrm{[\alpha/Fe]} = 0$ (top panel), and $\mathrm{[Fe/H]} = +0.3$ and $\mathrm{[\alpha/Fe]} = 0$ (bottom panel). The black point corresponds to the fitted value of the RGB bump in Baade's window, assuming a mean age for the stars of 10 Gyr, and a mean solar metallicity.}
\label{fig:Baade_RGBB}
\end{figure}

\subsection{Estimating the age of the dominant population of stars}
\label{sec:fit_ages}

As a proof of concept, in this Section we derive age estimates for the bulge field in Baade's window, showing the importance of a clean astrometric foreground removal to infer the overall distribution of ages of stars in the Milky Way bulge. In Fig. \ref{fig:Baade_Mi0} we show the distribution of absolute magnitudes for sources with $(g-i)_0 \geq 0.8$, to exclude the contamination from blue stars and from the MSTO. Following \citet{Surot19b}, we fitted the distribution with the sum of a fourth-order polynomial (for the background density) and three Gaussian distributions. While the third Gaussian is not needed to explain the observed trend, the RC (purple curve) and the red giant branch (RGB) bump (blue curve) are clearly visible in the luminosity function. We find that the RC has a mean magnitude $M_{i0} = 0.61$, with a standard deviation of $0.31$. The RGB bump is instead observed at $M_{i0} = 1.34$, with a standard deviation of $0.25$. 

\begin{table}
\centering
\caption{Best fit parameters for equation \eqref{eq:Delta_t} for three different combinations of $[\mathrm{Fe/H}]$ and $[\mathrm{\alpha/Fe}]$ explored in this work.}
\label{tab:Deltat}
\begin{tabular}{c c c c}
	\hline\hline
    $[\mathrm{Fe/H}]$ & $[\mathrm{\alpha/Fe}]$ & $a \ [\mathrm{Gyr}^{-1}]$ & $b$\\
    \hline
    $0$ & $0$ & $0.039 \pm 0.004$ & $0.17 \pm 0.04$ \\
    $-0.1$ & $+0.1$ & $0.041 \pm 0.004$ & $0.15 \pm 0.04$ \\
    $-0.2$ & $+0.2$ & $0.040 \pm 0.003$ & $0.14 \pm 0.03$ \\
    \hline
\end{tabular}
\end{table}

Since the absolute magnitudes are not calibrated, but they are just computed assuming a mean distance for all the stars corresponding to the mean distance of the RC stars in the field \citep{Johnson22}, we cannot directly compare their location to the theoretical values predicted by the MIST isochrones. 
For this reason, here we discuss the use of the distance from the RGB bump to the RC as an age indicator for the dominant stellar population in a given bulge field. We derive the location of the RGB bump and of the RC in MIST isochrones of different ages and metallicities, by visually inspecting the HR diagrams. Fig. \ref{fig:Delta_age_fit} shows the difference of the absolute magnitude in the $i$ band for the RGB bump and RC, $\Delta_\mathrm{RGBB-RC}$, as a function of the age of the population. The three chosen values of $[\mathrm{Fe/H}]$ and $[\mathrm{\alpha/Fe}]$ correspond to the adopted values in Baade's window, and in the fields at $(\ell,b) = (0^\circ, -6^\circ)$ and $(\ell,b) = (+5^\circ, -8^\circ)$ explored in Appendix \ref{app:age_highb}. The coloured lines correspond to linear fits to the data points:
\begin{equation}
    \label{eq:Delta_t}
    \Delta_\mathrm{RGBB-RC} = at + b \ ,
\end{equation}
where $t$ is the age of the isochrone in Gyr. The fitted parameters $a$ and $b$, together with their uncertainties, are provided in Table \ref{tab:Deltat} for three values of metallicities and abundances. The three fits are consistent with each other, even if it is possible to observe a trend at a given age: isochrones with lower values of $[\mathrm{Fe/H}]$ (and higher values of $[\mathrm{\alpha/Fe}]$) show systematically lower values of the $i$-band magnitude distance between the RGB bump and the RC.

We can now use the relation between the age and the magnitude difference between the RGB bump and the RC to estimate the age of the dominant stellar population in Baade's window. By inverting equation \eqref{eq:Delta_t}, given the fitted positions of the RGB bump and RC, we can estimate the ages as:
\begin{equation}
    \label{eq:t_Delta}
    t = \frac{\Delta_\mathrm{RGBB-RC} - b}{a}.
\end{equation}
The corresponding uncertainty on the age can then be estimated using error propagation:
\begin{equation}
    \label{eq:t_unc}
    \sigma_t = \sqrt{ \frac{(\Delta_\mathrm{RGBB-RC} - b)^2 \sigma_a^2}{a^4} + \frac{\sigma_b^2}{a^2} + \frac{\sigma_\Delta^2}{a^2} } \ ,
\end{equation}
where $\sigma_a$ and $\sigma_b$ are the uncertainties on the parameters $a$ and $b$, provided in Table \ref{tab:Deltat}, and $\sigma_\Delta = \sqrt{\sigma_\mathrm{RGBB}^2 + \sigma_\mathrm{RC}^2}$ is the combined uncertainty on the locations of the RGB bump and RC, which we assume $\sigma_\mathrm{RGBB} = \sigma_\mathrm{RC} = 0.1$ mag.

Using equations \eqref{eq:t_Delta} and \eqref{eq:t_unc}, and the values of the RGB bump and RC derived from the Gaussian fits, we obtain $\Delta_\mathrm{RGBB-RC} = 0.73 \pm 0.14$, and therefore a mean age for stars in Baade's window of $(14.13 \pm 3.92)$ Gyr. This value is shown with a black dot in Fig. \ref{fig:Delta_age_fit}. This age estimate, despite the large uncertainties, points to the picture of a predominantly old bulge, and is consistent with other measurements in the same field \citep[e.g.][]{Zoccali03, Clarkson08, Schultheis17}. 

This is further confirmed by Fig. \ref{fig:Baade_RGBB}, where in the top panel we show a selection of MIST isochrones for $\mathrm{[Fe/H]} = 0$ and $\mathrm{[\alpha/Fe]} = 0$. If we make the assumption that the mean age of the stellar population of the field is 10 Gyr, and that it has the same metallicity and alpha abundances as the Sun, then the fitted standard deviation of $0.25$ of the Gaussian distribution corresponding to the RGB bump in Baade's window can give us an approximate estimate of the minimum age of the stars in the field. By looking at the theoretical isochrones, we see how the observed uncertainty is marginally consistent with the location of the RGB bump for a population of $6$ Gyr, while stellar ages $\leq 5$ Gyr are excluded at $1$ sigma level. The bottom panel of Fig. \ref{fig:Baade_RGBB} shows instead MIST isochrones for $\mathrm{[Fe/H]} = +0.3$ and $\mathrm{[\alpha/Fe]} = 0$. If we now assume that the bulk of the population is best described by the MIST isochrone with $t = 10$ Gyr and $\mathrm{[Fe/H]} = 0$, then we see how a variation of $+0.3$ dex in metallicity extends the range of ages compatible with the observations, and only ages $< 4$ Gyr can be excluded at $1$ sigma level.

Because of the complex morphology of the Galactic bulge, this simplistic approach is not suited for fields at higher latitudes, where the X-shaped structure results into the overlapping along the line of sight of stellar populations at different distances. In Appendix \ref{app:age_highb} we further discuss this, showing an example from a more complex field at $(\ell, b) = (0^\circ, -6^\circ)$.

\section{Summary and Conclusions} \label{sec:summary}

In this analysis, we have demonstrated the power of using precise {\Gaia} EDR3 astrometry to remove the vast majority of foreground star contamination brighter than the MSTO, and produce clean BDBS optical CMDs for stars in the bulge. No photometric information was used to isolate the bulge stars from the foreground population. We analysed 127 different fields in the southern Galactic bulge, spanning $\ell \leq 9.5^\circ$ and $-9.5^\circ \leq b \leq -2.5^\circ$. {\Gaia} EDR3 parallaxes were used to remove obvious foreground stars with precise parallaxes consistent with the stars being within a few kpc from the Sun. Additionally, we show how {\Gaia} EDR3 proper motions can be further used to separate foreground stars from the bulge population. We used a GMM algorithm to fit two bivariate Gaussian distributions to the distribution of proper motions in Galactic longitude and latitude, which we then employed to assign to each star a probability to belong to the bulge according to its measured astrometry. We can summarize the main results of this paper as follows:

\begin{itemize}
    \item We produced extinction-corrected and foreground-cleaned bulge CMDs for $127$ individual fields in the southern Galactic bulge. The astrometric cleaning procedure removes the majority of the blue stars from each field, while retaining the population of red giants.

    \item The contamination from foreground stars with proper motions consistent with the bulge population in the CMDs is, at maximum, at a level of $\sim 20\%$ near the plane, with values $\sim 10\%$ at higher absolute values of the Galactic latitude.
    
    \item The overall ratio of foreground to background stars is found to be maximum for the fields at lower absolute Galactic latitudes, where it can be as high as $\sim 50\%$. This value decreases further away from the plane, reaching $\sim 25\%$.
    
    \item We produced kinematic transverse maps, recovering the expected kinematic patterns caused by the orientation of the bar and the morphological structure of the Milky Way bulge. The proper motion dispersions peak at $\sim 2.8$ mas yr$^{-1}$ near the plane because of the potential well of the Milky Way bulge, and decrease with increasing distance from the plane. The asymmetry of the dispersions with Galactic longitude is caused by the orientation of the bar. The radially-aligned quadrupole pattern observed in the proper motion correlation is a direct consequence of the X-shaped structure of the bulge.
    
    \item We quantified the fraction of blue to red bulge stars brighter than the MSTO, finding this to be around $3\%$ closer to the plane, and increasing to $\sim 20\%$ for the fields further away from the plane. This value is driven by the prominence of the RC in the CMDs of the fields at lower latitudes, and it should not be interpreted as an estimate of the residual contamination from stars belonging to the stellar disk.
    
    \item The population of blue stars visible especially in the fields closer to the Galactic plane is likely to be a residual foreground population. This particularly evident when investigating the kinematics of the red and blue stars in each field, and when applying more severe cuts in proper motions to isolate a pure bulge sample.
    
    \item The application of {\Gaia} EDR3 astrometric cleaning clearly shows that no widespread population of young He-burning blue loop stars is present across the bulge, and places strong limits on stars younger than 2 Gyr.

    \item We have derived a tight linear relation between the difference of the $i$-band magnitudes of the RGB bump and of the RC as a function of age, constructed from a set of MIST isochrones for different values of $[\mathrm{Fe/H}]$ and $[\mathrm{\alpha/Fe}]$. We have used this relation to derive a mean age of $(14.13 \pm 3.92)$ Gyr for stars in Baade's window. If we assume that the mean age of the population is 10 Gyr \citep{Schultheis17}, then contribution from stars with ages $< 6$ Gyr ($< 4$ Gyr) can be excluded at $1$ sigma level for a mean metallicity $[\mathrm{Fe/H}] = 0$ ($[\mathrm{Fe/H}] = +0.3$). The application of this method to fields at higher Galactic latitudes, where projection effects due to the morphology of the Galactic bulge become important, requires further work on the modelling of the populations.
    
\end{itemize}

Most of the interesting star formation history of the bulge, and the origin of the bar, will very likely require ages of $1-2$ Gyr precision for stars older than $5$ Gyr. These efforts offer the possibility of constraining whether the bar formed substantially later than the thick disk, and potentially whether the halo population is older still. With advanced kinematic models of the bulge and the derivation of photometric metallicities at the MSTO level \citep[e.g.][]{Brown+10}, one can envision some significant advances in age constraints, especially in the outer bulge fields where crowding and reddening are less severe. Future improvements in the precision of parallaxes and proper motions may be of great value in this endeavour. The extended observational baseline of future {\Gaia} data releases will result into a more precise and accurate determination of the astrometry of the stars, allowing us to refine the cleaning procedure and to lower the contamination from foreground objects. In particular, {\Gaia} DR4 (DR5) will be based on $5.5$ years ($10$ years) of data, and parallax precisions will improve by a factor of $\sim 1.33$ ($\sim 1.90$) with respect to {\Gaia} EDR3. For proper motions instead, the improvement will be $\sim 2.4$ and $\sim 7.1$ for {\Gaia} DR4 and {\Gaia} DR5, respectively, compared to {\Gaia} EDR3\footnote{\url{https://www.cosmos.esa.int/web/gaia/science-performance}}. Looking at the future, the proposed space mission {\Gaia}NIR, an all-sky near-infrared survey, will allow probing the central region of our Galaxy, providing photometry and astrometry for more than $10$ billion stars \citep{Hobbs16, Hog21}. Proper motions (parallaxes) from {\Gaia}NIR will be ~$10-20$ ($\sqrt{2}$) times more precise than the final data release of {\Gaia}, thanks to the improved baseline of $20$ years \citep{Hobbs19}. {\Gaia} DR5 will also provide radial velocities for all sources brighter than the $16$th magnitude in the Radial-Velocity Spectrometer (RVS) band. In addition, the advent of multi-object spectroscopic facilities such as MOONS \citep{MOONS} and 4MOST \citep{4MOST}, will complement {\Gaia} observations, providing spectra for million of stars in the bulge \citep{Chiappini+19}. The combined three-dimensional velocity will be essential to produce a precise foreground subtraction for the brightest stars, allowing the study of bulge CMDs to an unprecedented level of details.

\vspace{5mm}
\begin{acknowledgements}

The authors thank the anonymous referee for their comments, which greatly improved the quality of this manuscript.
T.M.\ thanks M. Rejkuba for interesting discussions.
T.M.\ acknowledges an ESO Fellowship. 
M.J.\ was supported primarily by the Lasker Data Science Fellowship, awarded by the Space Telescope Science Institute. M.J.\ further acknowledges the Kavli Institute for Theoretical Physics at the University of California, Santa Barbara, whose collaborative residency program TRANSTAR21 was supported by the National Science Foundation under Grant No. NSF PHY-1748958.
A.M.K. acknowledges support from grant AST-2009836 from the National Science Foundation. 
A.J.K.-H. gratefully acknowledges funding by the Deutsche Forschungsgemeinschaft (DFG, German Research Foundation) -- Project-ID 138713538 -- SFB 881 (``The Milky Way System''), subprojects A03, A05, A11. 

Data used in this paper comes from the Blanco DECam Survey Collaboration. This project used data obtained with the Dark Energy Camera (DECam), which was constructed by the Dark Energy Survey (DES) collaboration. Funding for the DES Projects has been provided by the U.S. Department of Energy, the U.S. National Science Foundation, the Ministry of Science and Education of Spain, the Science and Technology Facilities Council of the United Kingdom, the Higher Education Funding Council for England, the National Center for Supercomputing Applications at the University of Illinois at Urbana-Champaign, the Kavli Institute of Cosmological Physics at the University of Chicago, the Center for Cosmology and Astro-Particle Physics at the Ohio State University, the Mitchell Institute for Fundamental Physics and Astronomy at Texas A\&M University, Financiadora de Estudos e Projetos, Funda\c{c}\~{o} Carlos Chagas Filho de Amparo \'a Pesquisa do Estado do Rio de Janeiro, Conselho Nacional de Desenvolvimento Cient\'{\i}fico e Tecnol\'ogico and the Minist\'erio da Ci\^encia, Tecnologia e Inovac\~{a}o, the Deutsche Forschungsgemeinschaft, and the Collaborating Institutions in the Dark Energy Survey. The Collaborating Institutions are Argonne National Laboratory, the University of California at Santa Cruz, the University of Cambridge, Centro de Investigaciones En\'ergeticas, Medioambientales y Tecnol\'ogicas-Madrid, the University of Chicago, University College London, the DES-Brazil Consortium, the University of Edinburgh, the Eidgen\"ossische Technische Hochschule (ETH) Z\"urich, Fermi National Accelerator Laboratory, the University of Illinois at Urbana-Champaign, the Institut de Ci\'oncies de l'Espai (IEEC/CSIC), the Institut de F\'{\i}sica d’Altes Energies, Lawrence Berkeley National Laboratory, the Ludwig-Maximilians Universit\"at M\"unchen and the associated Excellence Cluster Universe, the University of Michigan, the National Optical Astronomy Observatory, the University of Nottingham, the Ohio State University, the OzDES Membership Consortium the University of Pennsylvania, the University of Portsmouth, SLAC National Accelerator Laboratory, Stanford University, the University of Sussex, and Texas A\&M University. Based on observations at Cerro Tololo Inter-American Observatory (2013A-0529; 2014A-0480; PI: Rich), National Optical Astronomy Observatory, which is operated by the Association of Universities for Research in Astronomy (AURA) under a cooperative agreement with the National Science Foundation. 

This work has made use of data from the European Space Agency (ESA) mission {\it Gaia} (\url{https://www.cosmos.esa.int/gaia}), processed by the {\it Gaia} Data Processing and Analysis Consortium (DPAC, \url{https://www.cosmos.esa.int/web/gaia/dpac/consortium}). Funding for the DPAC has been provided by national institutions, in particular the institutions participating in the {\it Gaia} Multilateral Agreement.

\emph{Software used:} numpy \citep{numpy},
          matplotlib \citep{matplotlib},
          scipy \citep{scipy},
          astropy \citep{astropy13, astropy18},
          scikit-learn \citep{scikit-learn}, 
          pomegranate \citep{Schreiber17},
          astroML \citep{astroML},
          TOPCAT \citep{topcat1, topcat2}

\end{acknowledgements}

\bibliography{references}{}
\bibliographystyle{aa}


\begin{appendix} 

\section{Choosing the number of components for the Gaussian Mixture Model}
\label{app:GMM}

\begin{figure}[]
\centering
\includegraphics[width=\columnwidth]{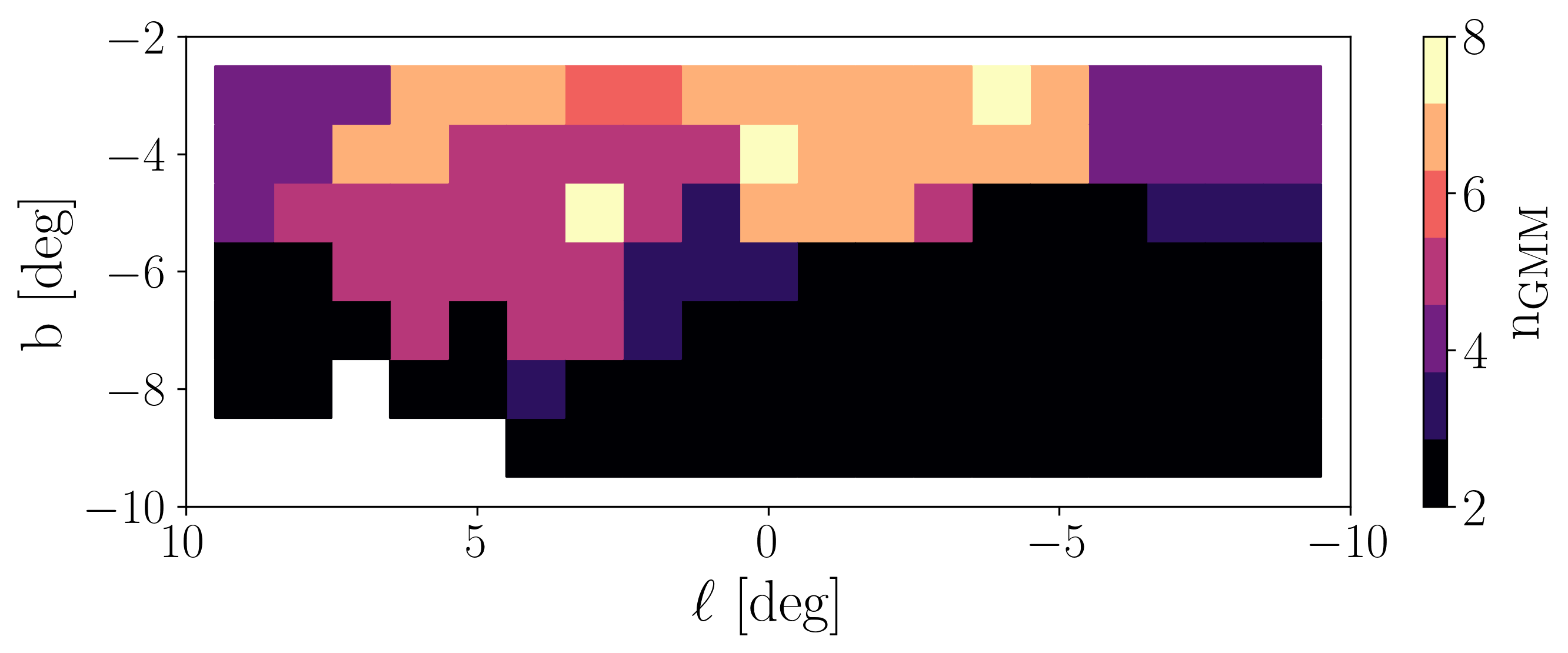}
\caption{Number of Gaussian components in the GMM minimizing the median BIC, for the bulge fields shown in Fig. \ref{fig:bulge_field}.}
\label{fig:GMM_n}
\end{figure}

We use the Bayesian information criterion (BIC) to select the optimal number of Gaussian components for the multivariate Gaussian distribution according to the GMM in each southern Galactic bulge field. Minimizing the BIC is a heuristic approach designed to select the model which best describes the data (maximizes the likelihood probability), but at the same time penalizes models with too many parameters (the total number of Gaussian distributions in this case), which might lead to overfitting. For each field, we remove $\varpi$-foreground stars using {\Gaia} EDR3 parallaxes, and then we apply the GMM to the {\Gaia} EDR3 proper motions, following the procedure outlined in Section \ref{sec:bulge_fields}. We explore a number of Gaussian components, ranging from $1$ (a single Gaussian distribution) to $8$. One of the Gaussian is initialized to the proper motion values obtained for giant stars, as discussed in Section \ref{sec:bulge_fields}, while all the others are initialized to $(\mu_{\ell *}, \mu_b) = (0, 0)$ mas yr$^{-1}$. We train the GMM $20$ times for each chosen value of the number of Gaussian components $n_\mathrm{GMM}$, and we compute the BIC as the median of the individual values, to compensate for the effect of the intrinsic randomness of the GMM algorithm. We then select the number of Gaussian components which minimizes the value of the median BIC. The result is shown in Fig. \ref{fig:GMM_n}, for the Bulge fields discussed in Section \ref{sec:bulge_fields}. This plot shows that $n_\mathrm{GMM} = 2$ for fields further away from the Galactic plane, and increases to higher values towards $b = 0$. The value of $n_\mathrm{GMM}$ is maximum near the plane, with $n_\mathrm{GMM}=8$ along the bulge minor axis. We also observe an asymmetry at positive values of Galactic longitude due to the presence of the near-side of the Galactic bar. We note that, even for the fields in which $n_\mathrm{GMM}=8$, the parameters of the Gaussian associated to the bulge stars (the red curve in Fig. \ref{fig:Baade_pms_GMM}) do not depend on the number of Gaussian components for the GMM, which affect only the $\mu$-foreground distributions at $\mu_{\ell *} \sim 0$ mas yr$^{-1}$. For this reason, in this work we choose to fit only two Gaussian components to the $\varpi$-background stars in the fields, to clearly isolate bulge and foreground stars. We refer to a future work the detailed investigation of the multiple population in proper motions for some fields, which might point towards the use of the precise {\Gaia} EDR3 proper motions to decompose the density field and identify not only bulge stars, but also differentiate between other stellar populations. Finally, we want to stress that $n_\mathrm{GMM}$ is not a precise determination of the exact number of individual populations in each field. A large value should be interpreted as an indication that a simple decomposition between bulge and foreground stars might be too simplistic, and it points to a higher complexity of the proper motion distribution in the field.

\section{Selecting the cleanest bulge CMDs}
\label{app:cleanest}

\begin{figure*}[]
\centering
\includegraphics[width=0.95\textwidth]{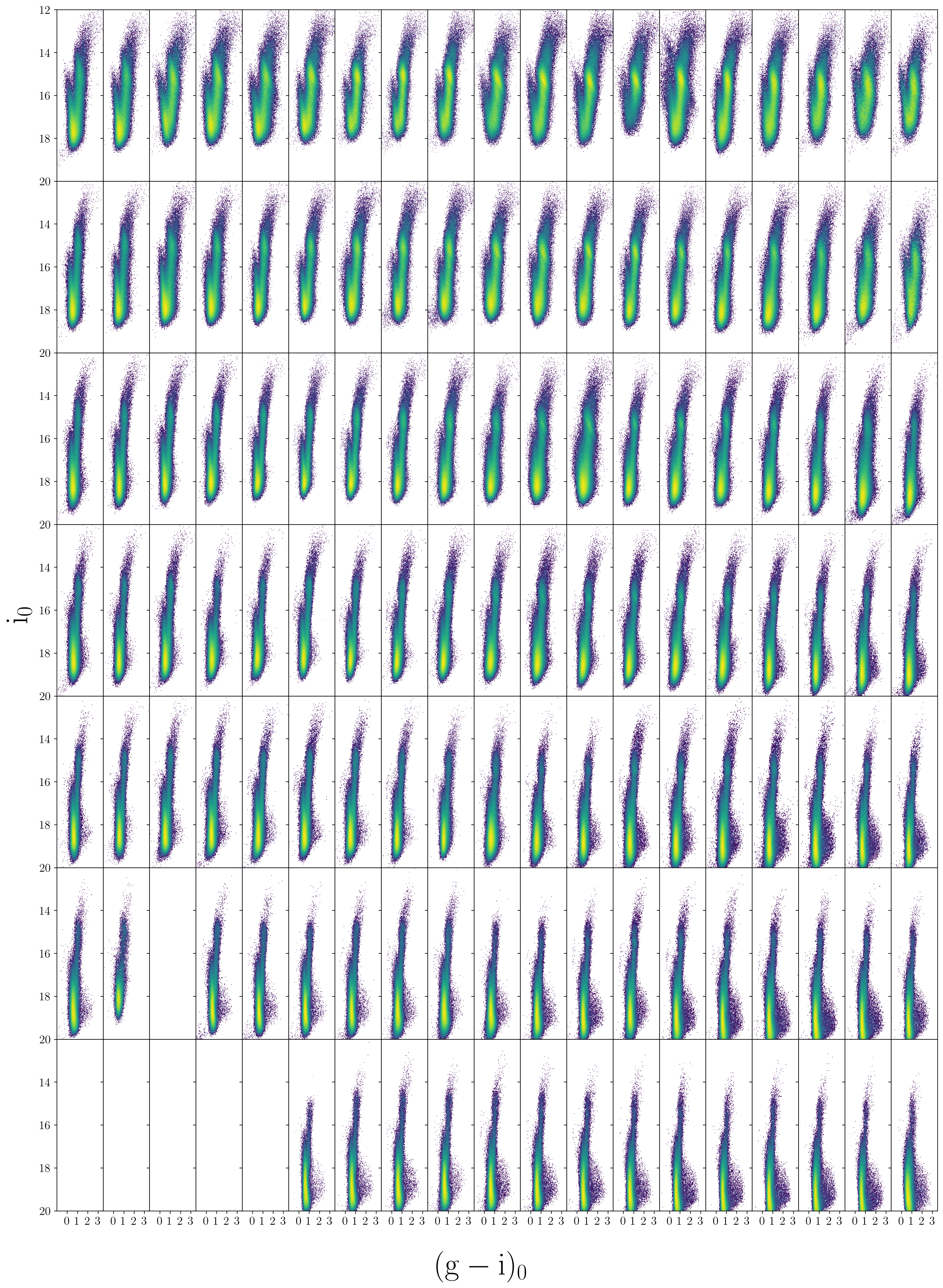}
\caption{Same as Fig. \ref{fig:CMDs_all_fields}, but considering only sources with $\mu_{\ell*} < -5$ mas yr$^{-1}$.}
\label{fig:CMDs_all_fields_cleanest}
\end{figure*}

\begin{figure}[]
\centering
\includegraphics[width=0.7\columnwidth]{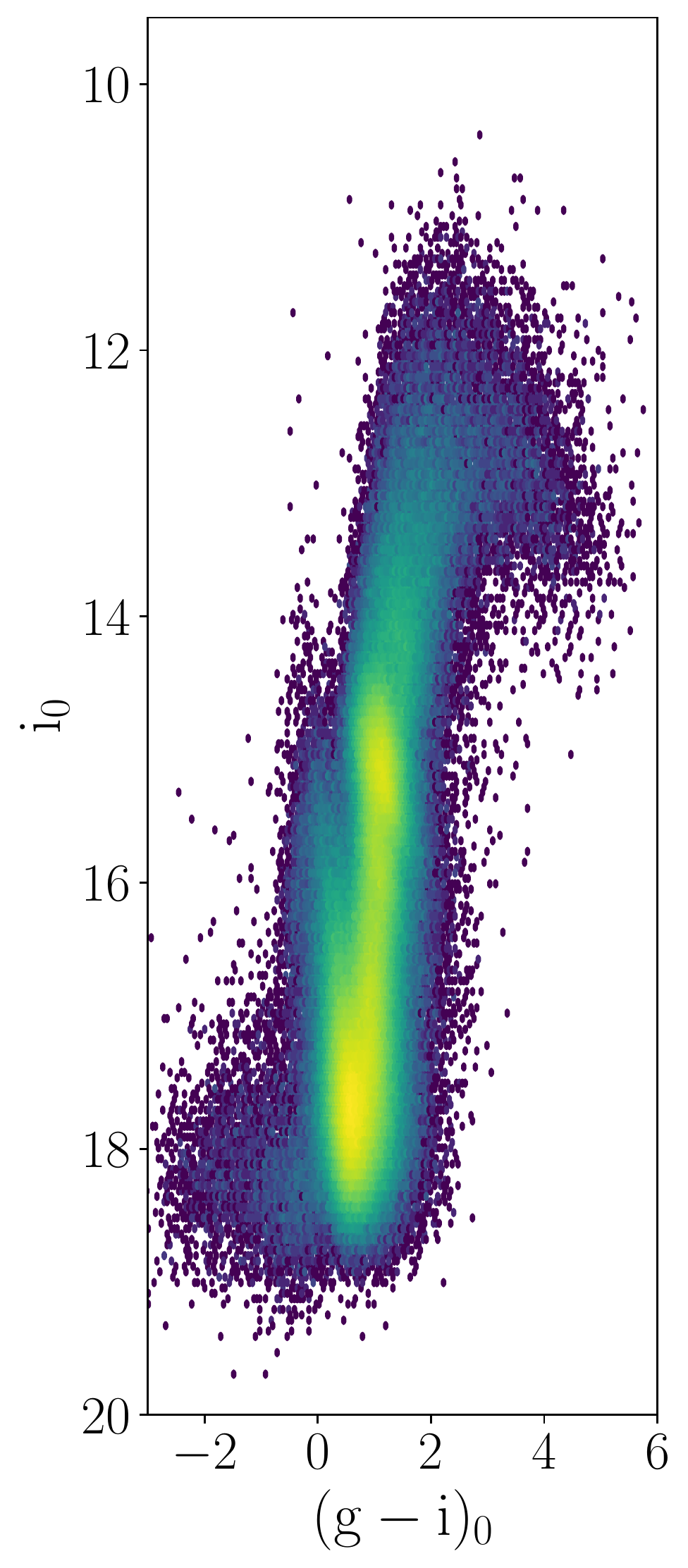}
\caption{CMD for all the $\varpi$-background stars in Baade's Window with $\mu_{\ell *} < -5$ mas yr$^{-1}$.}
\label{fig:Baade_pms_pure}
\end{figure}

In this appendix we further clean the bulge CMDs shown in Fig. \ref{fig:CMDs_all_fields}, reducing the completeness of the fields, but increasing their purity \citep[see also][]{Bernard18, Terry+20}. In Fig. \ref{fig:CMDs_all_fields_cleanest} we plot the bulge CMDs, selecting only sources with $\mu_{\ell *} < -5$ mas yr$^{-1}$. An individual CMD for stars in Baade's window is shown in Fig. \ref{fig:Baade_pms_pure}. As evident from Fig. \ref{fig:Baade_pms_GMM} and the first panel in Fig. \ref{fig:pm_moments}, this cut in proper motion removes the great majority of stars with proper motions consistent with belonging to the disk foreground population, and maximizes the contribution from the bulge stars. The resulting CMDs show a less prominent blue sequence, especially for the fields closer to the disk plane. This hints to the fact that the leftover blue population shown in Fig. \ref{fig:CMDs_all_fields} is likely to be a residual foreground contaminant to the bulge CMDs. This is due to two main factors: (i) the faintness of the sources prevents the use of {\Gaia} EDR3 parallaxes to determine their distances which would likely place those stars within a few kpc from the Sun, and (ii) the distribution in proper motions of disk and bulge stars show a significant overlap (as quantified by the first panel of Fig. \ref{fig:field_statistics}), and therefore a perfect separation between the two populations based on the astrometry alone is not possible.

\section{Age estimate for a field at higher Galactic latitudes}
\label{app:age_highb}

\begin{figure}[]
\centering
\includegraphics[width=0.92\columnwidth]{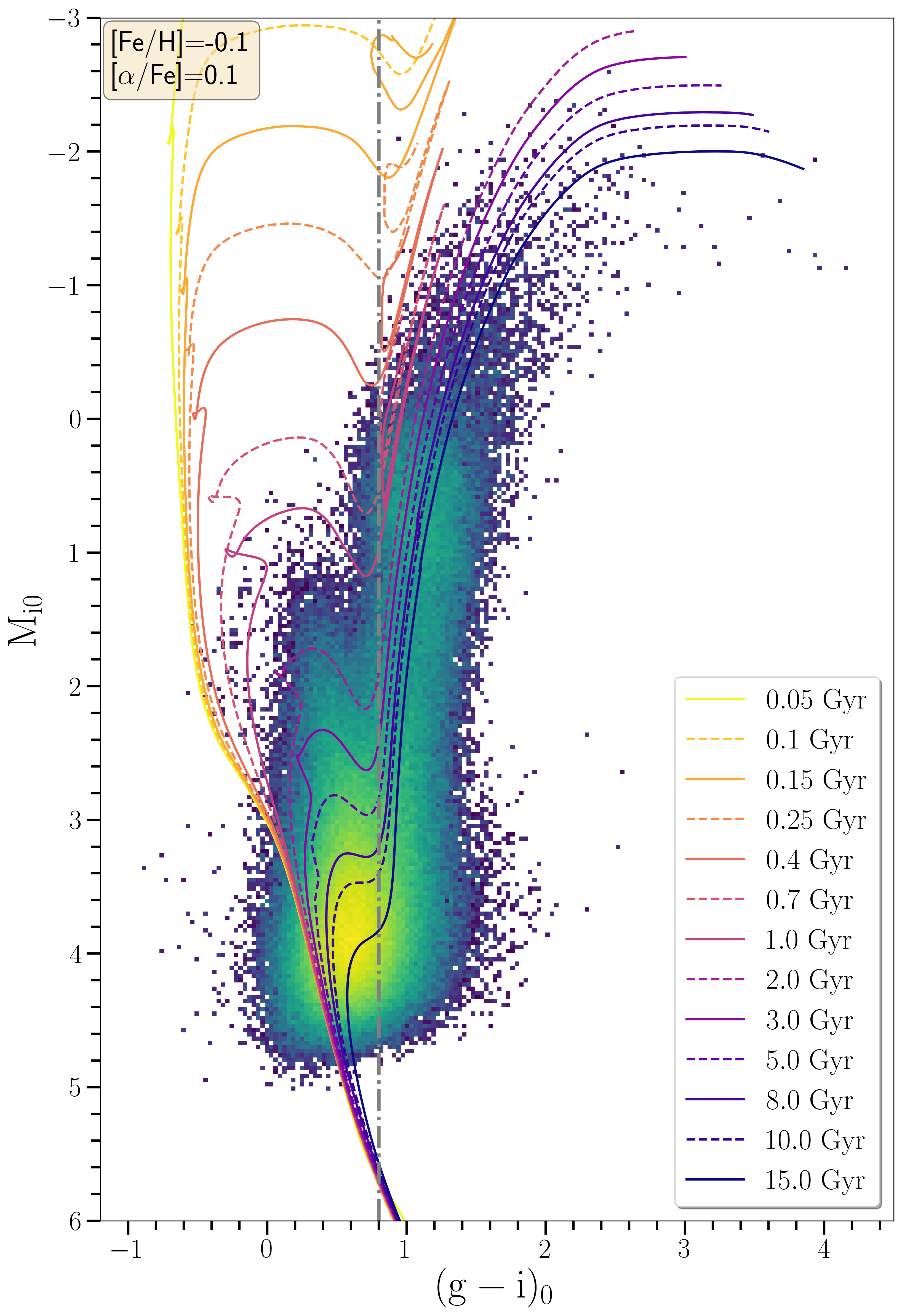}
\includegraphics[width=0.92\columnwidth]{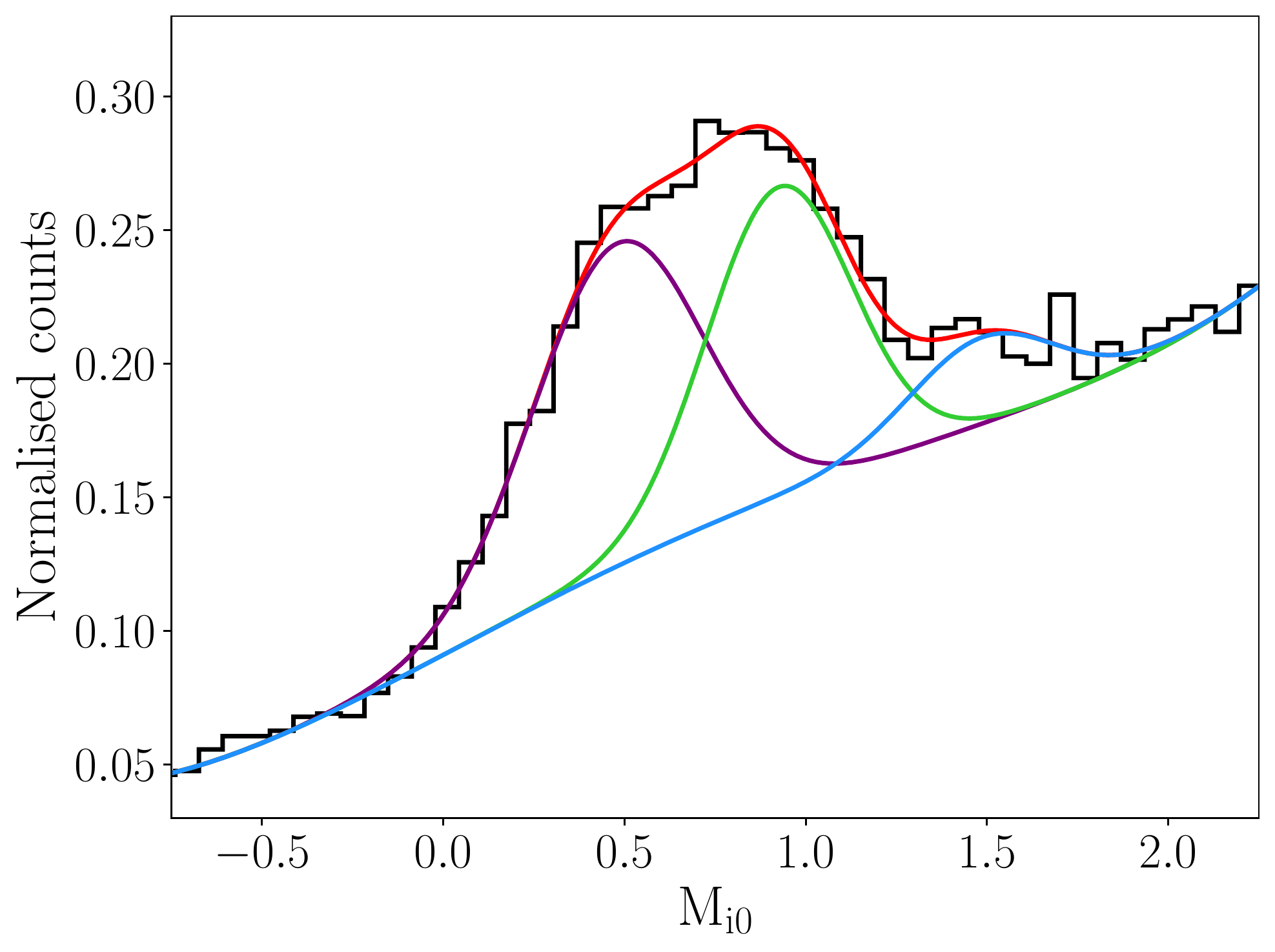}
\caption{\emph{Top:} Clean bulge de-reddened CMD for sources within $1^\circ$ from $(\ell, b) = (0^\circ, -6^\circ)$, with overplotted a selection of MIST isochrones with $[\mathrm{Fe}/\mathrm{H}] = -0.1$ and $[\mathrm{\alpha}/\mathrm{Fe}] = +0.1$. \emph{Bottom:} distribution absolute magnitudes for sources with $(g - i)_0 \geq 0.8$ (dash-dotted line in top panel), with Gaussian fits to the double RC (purple and green curve) and to the RGB bump (blue curve).}
\label{fig:iso_0-6}
\end{figure}

The top panel in Fig. \ref{fig:iso_0-6} shows the cleaned CMDs for the bulge field at $(\ell, b) = (0^\circ, -6^\circ)$. The field is chosen on the bulge minor axis, further away from the plane with the respect to Baade's window. We assume a distance of $8$ kpc to convert observed magnitudes into absolute ones, and we overplot MIST isochrones with $[\mathrm{Fe}/\mathrm{H}] = -0.1$ and $[\mathrm{\alpha}/\mathrm{Fe}] = +0.1$ \citep{Johnson22}. In this field, we can clearly rule out the presence of stars younger than 2 Gyr, and exclude a significant population of young stars with ages $\lesssim 5$ Gyr, which would have been apparent with a brighter MSTO. This is consistent with the results presented for Baade's window in Section \ref{sec:fit_ages}.

Following the same approach described in Section \ref{sec:fit_ages}, in the bottom panel of Fig. \ref{fig:iso_0-6} we show the luminosity function for stars with $(g - i)_0 \geq 0.8$. In this field, we clearly observe the double RC, which is caused by the X-shaped nature of the bulge, and the distance distribution of sources along this line of sight \citep[e.g.][]{Nataf10, McWilliam10, Gonzalez15, Lim21}. For the bright (faint) RC, we fitted a mean magnitude $M_{i0} = 0.48$ ($M_{i0} = 0.92$) and a standard deviation of $0.23$ ($0.20$). The RGB bump is instead observed at $M_{i0} = 1.49$, with a standard deviation of $0.20$. For this field, we try to fit $4$ Gaussian distributions to identify both components of the RGB bump, but we are not able due to its overlap with the RCs.

The application of the method introduced in Section \ref{sec:fit_ages}, using the distance between the RGB bump and the RC as a proxy for the age of the stars in the field, requires a precise (and unambiguous) detection of the two features in the observed CMD. A direct application of equation \eqref{eq:t_Delta} to the field at $(\ell, b) = (0^\circ, -6^\circ)$ would result in an unphysical age of $\sim 21$ Gyr, because of the large difference of $\sim 1$ mag between the RGB bump and the RC (see Fig. \ref{fig:Delta_age_fit}). This mismatch could be explained by the overlap between the faint RC and the bright component of the RGB bump, which is less populated compared to the RC \citep[e.g.][]{Renzini88, Nataf11}. 
To summarize, the presence of stars with different metallicities and different distances along the line of sight complicates the picture, and a detailed modelling of the complex three-dimensional morphology of the Galactic bulge is required before this method can be applied to all the bulge fields discussed in this work.

\end{appendix}
\end{document}